\documentclass[%
 aip,
 amsmath,amssymb,
 reprint,%
]{revtex4-1}

\usepackage{graphicx}
\usepackage{dcolumn}
\usepackage{bm}

\usepackage[utf8]{inputenc}
\usepackage[T1]{fontenc}
\usepackage{mathptmx}
\usepackage{etoolbox}
\usepackage{subcaption}
\usepackage{appendix}
\usepackage{booktabs}
\usepackage{placeins}

\newcommand{\be}{\begin{equation}}
\newcommand{\ee}{\end{equation}}
\newcommand{\bea}{\begin{eqnarray}}
\newcommand{\eea}{\end{eqnarray}}
\newcommand{\lb}{\label}

\newcommand{\RomanNumeralCaps}[1]

\newcommand{\bcalA}{\mbox{\boldmath ${\mathbb A}$}}

\makeatletter
\def\@email#1#2{%
 \endgroup
 \patchcmd{\titleblock@produce}
  {\frontmatter@RRAPformat}
  {\frontmatter@RRAPformat{\produce@RRAP{*#1\href{mailto:#2}{#2}}}\frontmatter@RRAPformat}
  {}{}
}%
\makeatother
\begin{document}

\preprint{AIP/123-QED}

\title{A Josephson-Anderson relation for drag in classical channel flows with streamwise periodicity: 
Effects of wall roughness}
\author{Samvit Kumar}
\author{Gregory L. Eyink}%
 \email{skumar67@jh.edu, eyink@jhu.edu}
\affiliation{ 
Department of Applied Mathematics and Statistics, Johns Hopkins University}%

\date{\today}

\begin{abstract}
The detailed Josephson-Anderson relation equates instantaneous work by pressure drop over any streamwise segment of a general channel and wall-normal flux of spanwise vorticity spatially integrated over that section. This relation was first derived by Huggins for quantum superfluids, but it holds also for internal flows of classical fluids and for external flows around solid bodies, corresponding there to relations of Burgers, Lighthill, Kambe, Howe and others. All of these prior results employ a background potential Euler flow with the same inflow/outflow as the physical flow, just as in Kelvin’s minimum energy theorem, so that the reference potential incorporates information about flow geometry. We here generalize the detailed Josephson-Anderson relation to streamwise periodic channels appropriate for numerical simulation of classical fluid turbulence. We show that the original Neumann b.c. used by Huggins for the background potential create an unphysical vortex sheet in a periodic channel, so that we substitute instead Dirichlet b.c. We show that the minimum energy theorem still holds and our new Josephson-Anderson relation again equates work by pressure drop instantaneously to integrated flux of spanwise vorticity. The result holds for both Newtonian and non-Newtonian fluids and for general curvilinear walls.  We illustrate our new formula with 
numerical results in a periodic channel flow with a single smooth bump, which reveals how vortex separation from the roughness element creates drag at each time instant. Drag and dissipation are thus related to vorticity structure and dynamics locally in space and time, with important applications to drag-reduction and to explanation of anomalous dissipation at high Reynolds numbers. 
\end{abstract}

\maketitle

\section{Introduction}

The modern paradigm\cite{packard1998role,varoquaux2015anderson}
for drag and dissipation in the theory of quantum superfluids arose from the work of \textcite{Josephson65} for superconductors and of \textcite{Anderson66} for neutral superfluids, who both noted a time-average relation between drops of voltage/pressure in flow through wires/channels and the cross-stream flux of quantized magnetic-flux/vortex lines. It was subsequently shown by \textcite{Huggins1970a} that a ``detailed Josephson-Anderson (JA) relation'' holds between instantaneous 
work by pressure drop and integrated flux of vorticity across the mass flux of the background potential associated to 
the ground-state quantum superflow. These results are the basis of contemporary solutions to the ``drag reduction'' problem 
in high-temperature superconductors, where, above some critical current, nucleation and motion of magnetic vortices 
creates an effective voltage drop and loss of superconductivity. The remedy is to introduce impurities and disorder 
to pin the vortices and prevent their cross-stream motion, thus restoring disspationless flow of electric current 
\cite{kwok2016vortices,glatz2020quest}. 

It was noted by \textcite{Anderson66} and by \textcite{Huggins1970a} that corresponding results hold 
for classical fluids described by the viscous Navier-Stokes equations. \textcite{eyink2008} pointed out that 
the time-average result had been invoked already by \textcite{taylor1932} for classical turbulent pipe flow and 
that an instantaneous relation between pressure gradients and vorticity flux at solid surfaces was derived 
by \textcite{lighthill1963}, both anticipating the results for quantum fluids. Subsequently, \textcite{eyink2021} showed that Huggins' detailed Josephson-Anderson relation holds also for external flows around solid bodies,
relating drag on the body instantaneously to the integrated flux of vorticity across the streamlines of the background potential flow. As reviewed by \textcite{biesheuvel2006force}, closely related instantaneous relations for drag in external flows of classical fluids had been previously derived by Burgers, Lighthill and others, especially \textcite{howe1995force}, and applied to both laminar and turbulent flow regimes. However, to our knowledge there has been no prior study applying the detailed relation of \textcite{Huggins1970a} to classical channel flows, either laminar or turbulent. Previous work of \textcite{huggins1994vortex},\textcite{eyink2008}, and \textcite{kumar2023flux} has investigated 
classical turbulent channel flow using only the time-averaged relation of \textcite{taylor1932}
and \textcite{Anderson66}, rather than the detailed relation which reveals the instantaneous 
connection between drag and vorticity dynamics. 

We shall show in this paper that the detailed Josephson-Anderson relation in the original form of 
\textcite{Huggins1970a} has in fact a significant flaw when applied to classical fluid turbulence. The
origin of the problem is Huggins' assumption that the channel inflow and outflow are pure potential, 
which is realistic for many superfluid applications where the quantum vortex tangle is strictly 
confined to some interior section of the channel. However, in applications to classical fluid turbulence 
this assumption is quite unrealistic as the outflow and very commonly the inflow as well consist 
of highly rotational flow. Furthermore, we shall see that Huggins' original derivation, when carried 
out with the streamwise periodic boundary conditions that are most common in numerical simulations, 
introduces a spurious vortex sheet into the reference ``potential'' flow. To avoid these 
serious difficulties we show here that it suffices to use instead a reference potential 
which matches only the mean mass flux of the physical flow and not the instantaneous inflow 
and outflow fields. We show nevertheless that the original derivation of \textcite{Huggins1970a} 
goes through with only minor modifications for this new choice of potential and yields again 
an instantaneous relation between work by pressure drop and spatially integrated vorticity flux. 
We then present a sample numerical application for turbulent channel flow with a single smooth bump
at modest Reynolds number, but sufficiently high that flow separation is observed with shedding 
of a rotational wake. In this flow we relate the instantaneous drag arising from both skin 
friction and pressure forces (form drag) to the vorticity flux from the boundary arising from separation. 
Our results thus reveal a deep unity to the origin of drag in both classical and quantum fluids. 

The results presented here build upon pioneering work of K.~R. Sreenivasan, who has made 
seminal contributions to turbulence in both quantum and classical fluids. In particular,
\textcite{bewley2008characterization} and \textcite{fonda2016sub,fonda2019reconnection}
developed the first experimental methods to visualize quantized vortices in a superfluid 
flow and to verify the reconnection dynamics which has been widely theorized to 
account for superfluid turbulent dissipation, going back to \textcite{feynman1955application}. 
We shall discuss below the relation of our results with such reconnection processes. In addition, 
\textcite{sreenivasan1987unified} and \textcite{sreenivasan1997persistent} have made 
fundamental contributions to the Reynolds-number scaling of turbulent wall-bounded flows, 
continuing in more recent works \cite{chen2021reynolds,chen2022law}. The persistent viscous 
effects identified by \textcite{sreenivasan1997persistent} make a very important contribution 
in particular to vorticity flux\cite{eyink2008,kumar2023flux} in wall-bounded flows, which is very relevant 
to our subject. Finally, the detailed Josephson-Anderson relation has direct applications to problems of polymer drag 
reduction studied by \textcite{sreenivasan2000onset} and turbulent energy dissipation rate 
studied in classic works of \textcite{sreenivasan1984scaling,sreenivasan1998update}, and \textcite{meneveau1991multifractal} which we discuss briefly below. A great legacy of Sreeni's research career is a strong interdiscplinary point 
of view and a search for general unifying principles, an example which we strive to emulate 
in this contribution to the Special Issue in honor of his 75th birthday.

\section{Prior Work of Huggins and Others}

In this section we very briefly review the detailed relation of Huggins, its derivation, and the closely
related results obtained by others for external flows. \textcite{Huggins1970a} considered a classical incompressible 
fluid at constant mass density $\rho$ and with kinematic viscosity $\nu$ subject to accelerations 
both from a conservative force $-\mathbf{\nabla} Q$ and from a non-conservative force $-\mathbf{f}$ 
satisfying $\bm{\nabla \times}\mathbf{f} \neq \bm{0}$, described by the incompressible Navier-Stokes 
equation written as 
\begin{align}
\partial_t\mathbf{u}&=\mathbf{u}\times \bm{\omega} -\nu\bm{\nabla\times\omega}-
\bm{\nabla}(p+|\mathbf{u}|^2/2+Q )-\mathbf{f}. \label{eq_mom}
\end{align}
A fundamental step made by \textcite{huggins1971dynamical,huggins1994vortex} was to rewrite the above 
momentum balance in the form 
\begin{align}
\partial_tu_i&=(1/2)\epsilon_{ijk}\Sigma_{jk}-\partial_i h,\label{JAsimple} \end{align} 
with anti-symmetric {\it vorticity flux tensor} 
\begin{align} \Sigma_{ij} &= u_i\omega_j-u_j\omega_i -\nu(\partial_i\omega_j-\partial_j\omega_i)-\epsilon_{ijk}f_k, 
\label{Sigma} \end{align} 
from vorticity advection, stretching, viscous diffusion and Magnus effect of the body force, and {\it total pressure} 
\begin{align} h&=p+|\mathbf{u}|^2/2+Q. \end{align}
including both the hydrostatic and the dynamic pressures. The tensor $\Sigma_{ij}$ represents the flux of the \textit{j}th vorticity component in the \textit{i}th 
coordinate direction. The latter interpretation is made clear by taking the curl of the momentum equation \eqref{eq_mom}, 
which yields a local conservation law for vector vorticity: 
\begin{align}
\partial_t\omega_j+\partial_i\Sigma_{ij} =0.
\end{align}
The equation \eqref{JAsimple} thus shows directly the connection between momentum balance and vorticity transport, and this equation is itself the most elementary version of the classical Josephson-Anderson relation. 

{ \begin{figure}
 \centering
\includegraphics[width=.49\textwidth]{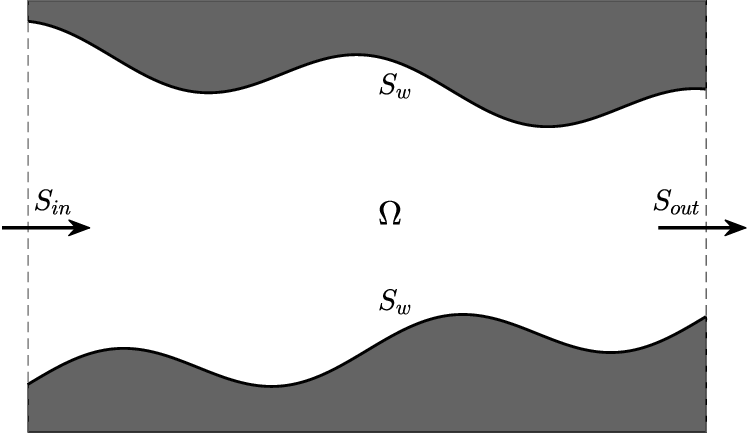}
\caption{Context of the detailed relation of \textcite{Huggins1970a}: Flow through a channel $\Omega$ 
with inflow surface $S_{in}$, outflow surface $S_{out}$, and sidewalls $S_w$.}
\label{scheme_huggins}
\end{figure}} 

To derive his detailed relation, \textcite{Huggins1970a} considered very general flows through pipes and channels, whose walls might be curved or bent, with rough or wavy surfaces, and with variable cross-sections. 
For example, a flow through an orifice in a wall is a classical application of the JA-relation in superfluids. 
See Figure~\ref{scheme_huggins} for the general situation. Huggins assumed given velocity at the inflow surface 
$S_{in}$ and at the outflow surface $S_{out}$ and stick boundary conditions at the sidewall $S_w$:  
\begin{align}
    \mathbf{u}|_{S_{in}}=\mathbf{u}_{in}, \qquad \mathbf{u}|_{S_{out}}=\mathbf{u}_{out}, \qquad  \mathbf{u}|_{S_{w}}=\bm{0}.
\end{align}
A key idea of \textcite{Huggins1970a} was then to compare the viscous rotational flow solving 
the incompressible Navier-Stokes equation \eqref{eq_mom} with an ideal incompressible potential flow 
$\mathbf{u}_\phi=\bm{\nabla}\phi$ solving the Euler equations with the same in-flow and out-flow: 
\begin{align}\label{bc_phi}
        \mathbf{n}\cdot\mathbf{u}_{\phi}|_{S_{in}}= \mathbf{n}\cdot\mathbf{u}_{in},& \qquad   
        \mathbf{n}\cdot\mathbf{u}_{\phi}|_{S_{out}}= \mathbf{n}\cdot\mathbf{u}_{out}, \qquad \\\nonumber
        & \mathbf{n}\cdot\mathbf{u}_{\phi}|_{S_{w}}=0.
\end{align}
where $\mathbf{n}$ is the unit normal at the boundary pointing into the fluid interior.
In superfluid applications this potential flow corresponds to the dissipation-less flow in the quantum 
ground state in the absence of any quantized vortex excitations. This is in fact the flow with the least energy 
among all incompressible flows with the boundary conditions \eqref{bc_phi} according to the  Kelvin minimum 
energy theorem \cite{kelvin1849vis,lamb1924hydrodynamics,batchelor_2000,wu2007vorticity}; see also below.   
The scalar potential $\phi$ solves the Laplace equation $\nabla^2\phi=0$ in the open flow domain $\Omega$ with 
Neumann boundary conditions supplied by \eqref{bc_phi} and is thus unique up to a spatial constant. In that case, the 
Euler dynamics reduce to the Bernoulli equation 
\begin{align}\label{bernoulli}
    \partial_t \phi+\frac{1}{2}|\mathbf{u}_{\phi}|^2+p_{\phi}+Q= c(t),
\end{align}
for a spatial constant $c(t),$ which yields the static Euler pressure $p_\phi$ and the total 
Euler pressure $h_\phi=p_{\phi}+|\mathbf{u}_{\phi}|^2/2+Q$ given the velocity potential $\phi.$
It is a direct consequence of \eqref{bernoulli} that the potential Euler solution experiences no mean drag,
since long-time averaging denoted by $\langle\cdot\rangle$ yields the relation 
\begin{equation}  \langle \bm{\nabla} h_\phi\rangle = \bm{0} \label{delamb} \end{equation} 
and thus mass flux occurs without any mean gradient of the total pressure. Likewise, 
in terms of the kinetic energy of the potential flow
\begin{equation}
E_{\phi}=(\rho/2)\int_{\Omega} |\mathbf{u}_{\phi}|^2dV,    \label{Ephi}  
\end{equation} 
one finds using the Bernoulli equation \eqref{bernoulli} that 
\begin{align}\label{e_phi2}
  \frac{dE_{\phi}}{dt} &= \rho \int_{\Omega}\mathbf{u}_{\phi}\cdot\bm{\nabla}(\partial_t \phi)\, dV= -\rho \int_{ \partial \Omega}  (\partial_t \phi) \mathbf{u}_{\phi}\cdot \mathbf{n}\, dA \nonumber \\
  &= \int_{S_{in}} h_\phi\, dJ-\int_{S_{out}} h_\phi \,dJ \ := \ {\mathcal W}_\phi,
\end{align} 
where $dJ=\rho \mathbf{u}_{\phi}\cdot d\mathbf{A}$ is the mass flux element along the potential flow
and where the last line defines the instantaneous rate of work ${\mathcal W}_\phi$ done by the potential 
pressure $h_\phi.$ As long as the inflow/outflow conditions remain bounded in time, then also 
$E_\phi$ remains bounded and long-time averaging yields
\begin{equation}  \langle {\mathcal W}_\phi\rangle = 0. \label{delamb2} \end{equation}
The relations \eqref{delamb},\eqref{delamb2} may be regarded as analogues of the ``d'Alembert paradox'' 
\citep{dalembert1749theoria,dalembert1768paradoxe} for potential fluid flows through pipes and channels. 

The detailed relation of \textcite{Huggins1970a} connected vortex motion further to energy balance. It was 
natural in Huggins' analysis to adopt as reference the potential flow $\mathbf{u}_\phi=\bm{\nabla}\phi$ which represents the superfluid velocity in the quantum ground state. Huggins thus decomposed the rate of work by 
the total pressure head 
\begin{align}     
\mathcal{W} =\int_{S_{in}}h\,dJ-\int_{S_{out}}h\,dJ 
\label{W-def} \end{align}
as $\mathcal{W}=\mathcal{W}_\phi+\mathcal{W}_\omega,$ where 
\begin{align}\label{Wom}
   \mathcal{W}_\omega &= \int_{S_{in}}h_{\omega} \,dJ -\int_{S_{out}}h_{\omega}\, dJ. 
\end{align}
is the rate of work done by the head of {\it total rotational pressure} $h_{\omega}=h-h_\phi.$ Because 
of \eqref{delamb2}, $\mathcal{W}_\omega$ represents the ``effective work'' which solely 
contributes to the long-time average. The main result of Huggins states 
that $\mathcal{W}_\omega$ is exactly equal to another quantity $\mathcal{T}$ that measures 
the flux of vorticity across the mass current of the background potential, given by the 
following equivalent expressions \newpage 
\begin{eqnarray}\label{Tdef}  
    \mathcal{T}
        &=& -\int_{\Omega} \rho\mathbf{u}_{\phi}\cdot (\mathbf{u}\times \bm{\omega}-\nu \bm{\nabla\times}\bm{\omega}-\mathbf{f})\, d V \cr
        &=& -\int dJ \int(\mathbf{u}\times \bm{\omega}-\nu \bm{\nabla\times}\bm{\omega}-\mathbf{f})\cdot d \ell \cr 
        &=&-\frac{1}{2}\int dJ \int \epsilon_{ijk}\Sigma_{ij}d\ell_k, 
\end{eqnarray}
where the line integrals are along streamlines of the potential flow. In fact, we shall see that 
$\mathcal{T}$ represents a transfer of kinetic energy from potential to rotational motions. The 
{\it detailed Josephson-Anderson relation} of \textcite{Huggins1970a} then states precisely the 
identity 
\begin{align}\label{JAdet}
   \mathcal{W}_\omega &= \mathcal{T}.
\end{align}
In other words, the effective rate of work done by the rotational pressure head is instantaneously related 
to the transverse motion of vortex lines across the potential flow. This relation is useful precisely because
$\mathcal{W}_\omega$ is the work contribution which is hard to understand and to compute, whereas $\mathcal{W}_\phi$
has transparent meaning and $\phi$ is computable at each time instant by standard solvers for the Laplace equation. 




 
Because we must generalize this result for classical turbulent channel flow, it is useful 
to reprise here the short proof. \textcite{Huggins1970a} obtained \eqref{JAdet} by deriving a 
complementary equation for the rotational fluid motions and by then considering the coupled energy 
balances for potential and rotational flows. The rotational velocity field defined by Huggins was 
$\mathbf{u}_{\omega}:=\mathbf{u}-\mathbf{u}_{\phi}$, which accounts for all vorticity 
in the flow. Its governing equations are easily obtained by subtracting the Euler equation for 
$\mathbf{u}_\phi$ from the Navier-Stokes \eqref{eq_mom}, yielding,
\begin{align}\label{u_omega}
    \partial_t\mathbf{u}_{\omega}&=\mathbf{u}\times \bm{\omega} -\nu\bm{\nabla}\times\bm{\omega}-\bm{\nabla} h_{\omega}\end{align}
where $h_{\omega}$ can be rewritten (up to a spatial constant) as 
\begin{align}   \label{hom-def}   
h_{\omega}&=h+\partial_t\phi=p_{\omega}/\rho+|\mathbf{u}_{\omega}|^2/2 +\mathbf{u}_{\omega}\cdot\mathbf{u}_{\phi}.
    \end{align}    
with $p_\omega=p-p_\phi$ and $\mathbf{u}_\omega$ satisfies the boundary conditions
\begin{align}\label{u_omegabc}
           \mathbf{u}_{\omega}|_{S_{in}}= (\mathbf{u}-\mathbf{u}_{\phi})|_{S_{in}}, & \qquad \mathbf{u}_{\omega}|_{S_{out}}=(\mathbf{u}-\mathbf{u}_{\phi})|_{S_{out}}, \\\nonumber
           & \mathbf{u}_{\omega}|_{S_{w}}=-\mathbf{u}_{\phi}|_{S_{w}}.
\end{align}
In particular, 
\begin{equation} \mathbf{n\cdot u}_{\omega}|_{S_{in}}= \mathbf{n\cdot u}_{\omega}|_{S_{out}}= \mathbf{n\cdot u}_{\omega}|_{S_{w}}=0.
\label{noflow-om} \end{equation}  
The latter render the potential velocity $\mathbf{u}_{\phi}$ 
and the rotational velocity $\mathbf{u}_{\omega}$ orthogonal, since their spatial $L^2$ inner product is 
\begin{align}
    \int_{\Omega} \mathbf{u}_{\phi}\cdot \mathbf{u}_{\omega} dV = \int_{\Omega} \bm{\nabla}\cdot(\phi\mathbf{u}_{\omega}) dV
    =-\int_{\partial \Omega} \phi \mathbf{u}_{\omega} \cdot \mathbf{n}\, dS=0.
\end{align}
This orthogonality is the essence of Kelvin's minimum energy theorem, since it 
implies that the total kinetic energy $E=(\rho/2)\int_{\Omega} |\mathbf{u}|^2dV$ in the channel is a sum of potential 
and rotational contributions, $E=E_\phi+E_\omega,$ with the kinetic energy of rotational motions given by 
$$ \qquad E_{\omega}=(\rho/2)\int_{\Omega} |\mathbf{u}_{\omega}|^2dV. $$
In that case, the minimum kinetic energy $E$ for all incompressible velocity fields $\mathbf{u}$ satisfying 
the b.c. \eqref{bc_phi} is obviously achieved with $\mathbf{u}_\omega=\mathbf{0}$ or $\mathbf{u}=
\mathbf{u}_\phi.$

From the above equations, \textcite{Huggins1970a} derived balance equations for $E_{\omega}$ and $E_{\phi}.$
Taking the dot product of \eqref{u_omega} with $\rho\mathbf{u}_{\omega}$ and integrating over the channel volume yields 
\begin{align}\label{e_omeg1}
        &\frac{dE_{\omega}}{dt}=\mathcal{T}-\mathcal{D} \nonumber \\
        &\hspace{20pt} + \int_{S_{in}} h_\omega \mathbf{u}_\omega\bm{\cdot}\mathbf{n}\, dA 
                       + \int_{S_{out}} h_\omega \mathbf{u}_\omega\bm{\cdot}\mathbf{n}\, dA,  
\end{align}         
so that the b.c. \eqref{noflow-om} give the final equation for $E_{\omega}$ as
\begin{align}\label{e_omega}
        &\frac{dE_{\omega}}{dt}=\mathcal{T}-\mathcal{D} 
\end{align}         
where $\mathcal{T}$ is given by \eqref{Tdef} and the total energy dissipation by non-conservative forces is 
given by 
\begin{align}         
\mathcal{D}&= \int_{\Omega} (\eta |\bm{\omega}|^2 +  \rho \mathbf{u}\cdot\mathbf{f})\, dV. \nonumber
\end{align}
with $\eta=\nu\rho$ the shear viscosity. The total energy satisfies of course the standard balance 
\begin{align}\label{e_tot}
    \frac{dE}{dt}&=\mathcal{W} -\mathcal{D}. \end{align} 
The equation for $E_{\phi}$ is 
then obtained simply by subtracting the equations \eqref{e_tot} and \eqref{e_omega}, yielding,
 \begin{align}\label{e_phi}
     \frac{dE_{\phi}}{dt}=\mathcal{W} -\mathcal{T}.
 \end{align}
 The two balance equations \eqref{e_omega},\eqref{e_phi} show that the work $\mathcal{W}$ 
 done by the pressure head goes entirely into potential flow energy, which is in turn transferred 
 by vortex motion through the term $\mathcal{T}$ into rotational flow energy, and then ultimately 
 disposed by the dissipation $\mathcal{D}$ due to viscosity and other non-ideal forces acting 
 on the rotational flow. As a final step, \textcite{Huggins1970a} then substituted the relation 
 \eqref{e_phi2} for $dE_\phi/dt$ into \eqref{e_phi} which, recalling the definition $\mathcal{W}_\omega:
 =\mathcal{W}-\mathcal{W}_\phi,$ yields directly the detailed Josephson-Anderson relation \eqref{JAdet}.

The previous results are very closely analogous to well-known results for external flows
around bodies in translational motion with velocity $-\mathbf{V}(t)$ or equivalently, by a change 
of reference frame, flows around bodies at rest with fluid velocity $\mathbf{V}(t)$ at infinity. 
We prefer to state the results in the latter body frame and we omit all proofs, referring 
to standard sources such as \textcite{batchelor_2000}, \textcite{lighthill1986informal}, 
\textcite{wu1981theory}, \textcite{eyink2021} and the review of \textcite{biesheuvel2006force}. The main object 
of interest here is the force acting on the fixed body $B$ 
$$ \mathbf{F}(t) = \int_{\partial B} (-P\mathbf{n}+\rho \bm{\tau}_w)\,dA  $$
with $P=\rho p$ the thermodynamic pressure and with $\bm{\tau}_w=\nu\bm{\omega\times}\mathbf{n}
=2\nu\mathbf{S}\bm{\cdot}\mathbf{n}$ the viscous skin friction. This force is of course related 
to fluid impulse $\mathbf{I}(t)$ by the well-known relation $\mathbf{F}(t)=-d\mathbf{I}/dt.$ In the 
special case of potential flow satisfying the no-penetration b.c. $\partial\phi/\partial n=0$ 
at the body surface $\partial B,$ the force is given by 
$$ \mathbf{F}_\phi(t) = - \int_{\partial B} P_\phi\mathbf{n}\,dA  $$
and the impulse by 
$$ \mathbf{I}_\phi(t) = -\rho \int_{\partial B} \phi\mathbf{n}\,dA  $$
once again related by $\mathbf{F}_\phi(t)=-d\mathbf{I}_\phi/dt.$ Thus,
$$ \langle \mathbf{F}_\phi\rangle =0 $$
which is the ``generalized d'Alembert paradox'' for bodies in non-uniform translational motion. 
As in the work of Huggins, \textcite{lighthill1986informal,lighthill1986fundamentals} 
and others \cite{biesheuvel2006force,eyink2021} have proposed to divide the flow into the background 
potential flow fields $\mathbf{u}_\phi,$ $p_\phi$ and the complementary rotational fields 
$\mathbf{u}_\omega=\mathbf{u}-\mathbf{u}_\phi,$ $p_\omega=p-p_\phi.$ The 
"effective force" imposed by rotational fluid motions is then 
$$ \mathbf{F}_\omega(t) = \int_{\partial B} (-P_\omega\mathbf{n}+\rho\bm{\tau}_w)\,dA  $$
and the impulse of the rotational flow is 
$$  \mathbf{I}_\omega(t)=\frac{1}{2} \left[\int_\Omega \mathbf{x}\bm{\times}\bm{\omega}(\mathbf{x},t)\,dV
     + \int_{\partial B} \mathbf{x}\bm{\times}(\mathbf{n}\bm{\times}\mathbf{u}_\omega(\mathbf{x},t))\,dA \right]$$
so that $\mathbf{F}_\omega(t)=-d\mathbf{I}_\omega/dt.$     
Note that $\bm{\Gamma}=\mathbf{n}\bm{\times}\mathbf{u}_\phi$ can be regarded as the strength vector of a surface vortex sheet 
of the potential flow $\mathbf{u}_\phi$ and since $\mathbf{n}\bm{\times}\mathbf{u}_\omega=-\mathbf{n}\bm{\times}\mathbf{u}_\phi$ on $\partial B$
$$  \mathbf{I}_\omega(t)=\frac{1}{2} \int_\Omega \mathbf{x}\bm{\times}\bm{\omega}_a(\mathbf{x},t)\,dV $$
where $\bm{\omega}_a$ is the so-called {\it additional vorticity}, with the surface vortex sheet 
removed. Note that generally $\langle\mathbf{F}_\omega(t)\rangle \neq \bm{0}$ because 
$\mathbf{I}_\omega(t)$ increases monotonically as the rotational wake grows in extent and 
its impulse is not bounded in time. 

In the present context of external flow, the quantity analogous to the rate of pressure work
\eqref{W-def} for channel flows is the power dissipated by the drag force: 
\begin{eqnarray}
    \mathcal{W}:= \mathbf{F}(t)\bm{\cdot} \mathbf{V}(t).  
\end{eqnarray}
The force decomposition $\mathbf{F}(t)=\mathbf{F}_\phi(t)+\mathbf{F}_\omega(t)$ immediately 
implies a corresponding decomposition of the dissipated power $\mathcal{W}(t)= 
\mathcal{W}_\phi(t)+\mathcal{W}_\omega(t).$ However, since impulse $\mathbf{I}_\phi(t)=\bcalA\bm{\cdot}\mathbf{V}(t)$ with $\bcalA$ a time-independent {\it added mass} tensor depending only on the shape of the body, it follows that $\mathcal{W}_\phi =\mathbf{F}_\phi(t)\bm{\cdot} \mathbf{V}(t)= \frac{d}{dt}\left(\frac{1}{2}\mathbf{V}(t)\bm{\cdot}\bcalA\bm{\cdot}\mathbf{V}(t)\right)$ and thus 
$$ \langle  \mathcal{W}_\phi\rangle =0. $$
Just as before, there is no time-average power dissipated by the potential drag force and all of the 
``effective dissipation'' arises from drag force due to rotational flow:
\begin{eqnarray}
    \mathcal{W}_\omega:= \mathbf{F}_\omega(t)\bm{\cdot} \mathbf{V}(t)  
\end{eqnarray}
It was shown by \textcite{eyink2021} that a detailed Josephson-Anderson relation holds for the latter,
of the same form as \eqref{JAdet}: 
$$ \mathcal{W}_\omega =\mathcal{T} $$
where $\mathcal{T}$ is given by exactly the same expression \eqref{Tdef}. Thus, power dissipated by 
drag on the body due to rotational fluid motions is given instantaneously by the space integral of the 
vorticity flux across the flowlines of the background potential. In fact, this result is just a special 
case of a more general result of \textcite{howe1995force} which applies to arbitrary rigid body motion
(translation and rotation) and which gives all force components, not only drag but also lateral 
forces such as lift.  

It was pointed out by \textcite{eyink2021} that the JA-relation should hold even in the limit of infinite 
Reynolds number, if spatial integration-by-parts is performed to rewrite the transfer term in 
\eqref{Tdef} instead as 
\be \mathcal{T}= -\rho \int_{\Omega}\mathbf{\nabla u_{\phi}}:\mathbf{u_{\omega}}\mathbf{u_{\omega}} \,dV
+ \rho\int_{\Omega} \mathbf{u_{\phi}}\cdot \mathbf{f} \, dV
+\rho\int_{\partial \Omega} \mathbf{u_{\phi}}\cdot \bm{\tau}_w \, dA.  
\lb{Tdef2} \ee
This mathematical 
conjecture of an infinite-$Re$ limit has been verified by \textcite{quan2022onsager} for the case 
of no body-force ($\mathbf{f}=\mathbf{0})$ in flow around a solid body, providing a new resolution 
of the famous paradox of  \textcite{dalembert1749theoria,dalembert1768paradoxe}
and connecting with the Onsager theory of ``ideal turbulence'' \citep{onsager1949statistical,eyink2006onsager,
eyink2024onsager}. To derive these conclusions for the limit $Re\to\infty$ it is crucial that
the reference potential flow velocity must be infinitely differentiable or $C^\infty,$ which 
could indeed be proved for external flow as long as the body surface is correspondingly smooth. 

\section{A New Detailed Relation for Streamwise Periodic Poiseuille Flows}

The previous developments reviewed above suggest that vorticity flux accounts for wall drag with 
great generality, in many incompressible fluid flows of practical and theoretical interest, and that 
the JA relation can provide a novel vorticity-based perspective on drag reduction. Unfortunately, 
a difficulty occurs in the straightforward application of the original relation of \textcite{Huggins1970a} 
to classical turbulent flows through pipes and channels. In that case, the fields $\mathbf{u}_{in}$ and 
$\mathbf{u}_{out}$ that appear in the boundary conditions \eqref{bc_phi} 
for the reference potential flow are both $x$-slices 
of a very complex and rough turbulent velocity field. This means that $\mathbf{u}_\phi$ is generally 
also spatially complex and rough, inheriting those properties from its boundary conditions. This poses a 
serious problem for mathematical analysis of the infinite-Reynolds limit\cite{quan2022onsager}, 
since the arguments involved depend crucially on the smoothness of the potential flow.  Furthermore, 
this non-smoothness of $\mathbf{u}_{in}$ and $\mathbf{u}_{out}$ makes more demanding 
the numerical computation of $\mathbf{u}_\phi.$ The corresponding problem does not appear 
in typical superfluid applications, since the vortex tangles in that case are generally 
confined well within the channel interior.

Another important issue is that numerical simulations of turbulent pipe and channel flows
in classical fluids very frequently employ periodic boundary conditions in the streamwise direction
as a computational convenience. The flow may be driven either with a fixed bulk velocity or 
as Poiseuille flow with a fixed pressure gradient. In the latter case, a standard choice is 
to use a non-periodic linear potential $Q_{nper}=-\gamma(t) x,$ where $x$ is taken as the streamwise 
direction and $\gamma(t)$ is the resultant streamwise gradient in the total pressure $h$, but with 
velocity $\mathbf{u}$ and static pressure $p$ both periodic. This situation is of the type considered 
by \textcite{Huggins1970a} but with the ends of the channel periodically joined so that $S_{in}=S_{out}.$ 
In this setting, the naive approach would be to mimic exactly the original derivation 
and take as reference field the Euler flow with potential $\phi$ solving Laplace's equation with Neumann 
boundary conditions \eqref{bc_phi}, without regard for the fact that $\mathbf{u}_{in}=\mathbf{u}_{out}.$
All of the analysis and results of \textcite{Huggins1970a} then carries over in this setting. However, there 
is a serious difficulty. The potential $\phi$ is uniquely specified (up to a spatial constant) 
by the Laplace problem with Neumann boundary conditions \eqref{bc_phi} and these conditions
guarantee that $\mathbf{n\cdot u}_{\phi,in}=\mathbf{n\cdot u}_{\phi,out}$  so that $\mathbf{n\cdot u}$ 
is $x$-periodic. However, in general the components of $\mathbf{u}_\phi$ perpendicular to $\mathbf{n}$ need not be periodic. 
In fact, any such discontinuity corresponds to a vortex sheet in $\mathbf{u}_\phi$ at $S_{in}=S_{out}$ 
with strength $\bm{\Gamma}=\mathbf{n\times}(\mathbf{u}_{\phi,out} - \mathbf{u}_{\phi,in}).$
Since the surface $S_{in}=S_{out}$ was arbitrarily chosen and any $x$-cross-section could be 
equally selected for the construction, this means that there is a vortex sheet in the interior
of the periodic domain and $\mathbf{u}_\phi$ is not truly potential.  

\begin{figure}
      \centering
      \begin{subfigure}[b]{0.235\textwidth}
         \centering
         \includegraphics[width=\textwidth]{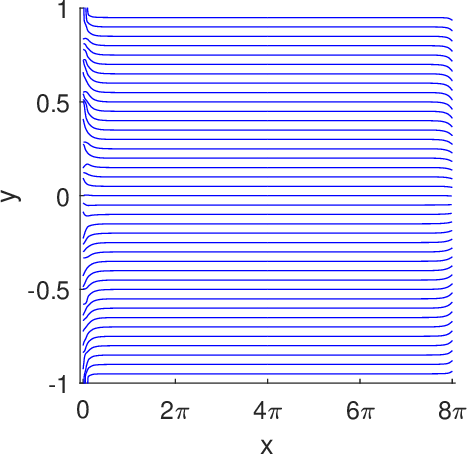}
         \caption{Huggins' potential}
         \label{pot_huggins}
         \end{subfigure}
         \hfill
     \begin{subfigure}[b]{0.235\textwidth}
         \centering
         \includegraphics[width=\textwidth]{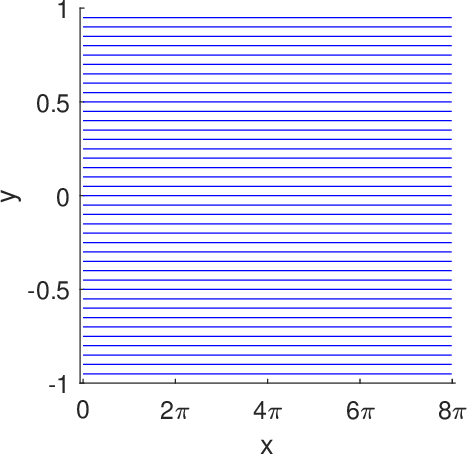}
        \caption{Modified potential}
         \label{pot_new}
     \end{subfigure}

\caption{Streamlines of the reference potential for a snapshot of streamwise-periodic 
turbulent channel flow}   \label{pot_lines}
\end{figure}

\begin{figure*}
\makebox[\textwidth][c]{\includegraphics[width=0.8\textwidth]{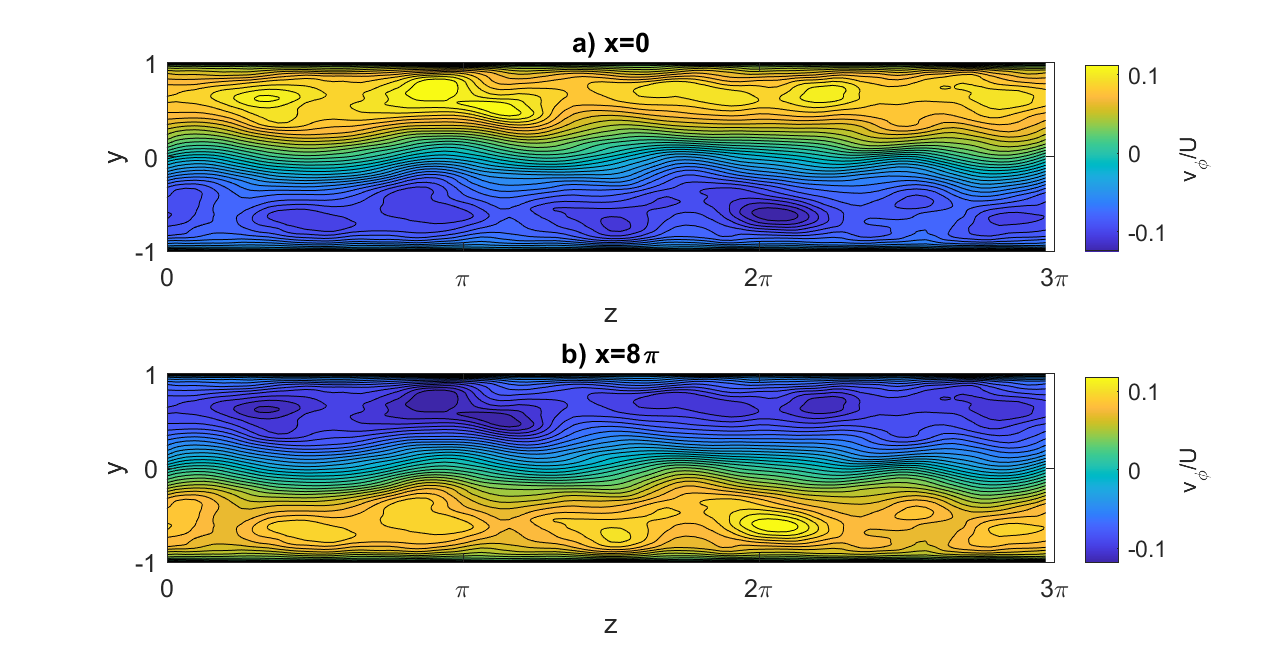}}
\caption{
Anti-periodic wall-normal component of Huggins' potential flow velocity for a snapshot of turbulent channel flow,
at inflow $x=0$ and outflow $x=8\pi.$}
\label{vphi_in_out.fig} 
\end{figure*}

Both of these problems can be illustrated in the case of turbulent Poiseuille flow through 
a smooth plane-parallel channel, using data from the Johns Hopkins turbulence database 
(JHTDB) \citep{jhtdb1,jhtdb_channel}
which hosts data from a numerical simulation at $Re_\tau=1000$ on a space domain $[0,8\pi]\times
[-1,1]\times [0,3\pi]$ with periodic b.c. in the streamwise $x$-direction and spanwise $z$-direction, but 
stick b.c. in the wall-normal $y$-direction. We have obtained Huggins' reference potential $\phi$ by solving 
numerically Laplace's equation with boundary conditions \eqref{bc_phi}, using a 2nd-order central-difference
scheme. The streamlines of this potential for one time snapshot from the database are plotted in panel (a) 
of Fig.~\ref{pot_lines} and show spatially irregular behavior near in-flow at $x=0$ and out-flow at $x=8\pi.$ 
The same irregularity is observed in the results for the wall-normal velocity component $v_\phi$ 
plotted in Fig.~\ref{vphi_in_out.fig} at in-flow and out-flow. Even more seriously, 
this velocity component can be seen to be streamwise anti-periodic as is also the spanwise component $w_\phi$
(see Supplementary Materials, \S I), both corresponding to a vortex sheet in $\mathbf{u}_\phi.$
Interestingly, however, after inertial adjustment over a length of order the channel half-width, 
the potential flow field closely resembles a plug flow with spatially constant velocity 
$\mathbf{u}_\phi=U\hat{\mathbf{x}},$ for $U$ the bulk flow velocity. The latter observation suggests 
that it might be possible in this case to use as reference flow the simple Euler solution 
$\mathbf{u}_\phi=U\hat{\mathbf{x}}$ with non-periodic potential $\phi=Ux$ which has constant values 
$\phi=0$ at $x=0$ and $\phi=8\pi U$ at $x=8\pi.$ This idea is readily verified.

 Motivated by this example, however, we show here that one may more generally 
derive a Josephson-Anderson relation for streamwise-periodic Poiseuille flows using a potential $\phi$
satisfying Dirichlet b.c. at the end sections and which is periodic plus a linear part. 
We can consider generalized pipe and channel flows with curved or rippled walls, but, for technical 
reasons explained below, we must assume that the flow domain extends over $x\in [-0.5L_x,0.5L_x]$ and 
$S_{in},$ $S_{out}$ are flat surfaces of constant $x$-value. For the case of channels we assume likewise 
a spanwise extent $z\in [-0.5L_z,0.5L_z]$, with periodic b.c. Finally, the flow is assumed driven by a 
non-periodic potential $Q_{nper}=-\gamma(t) x $,  with fluid velocity $\mathbf{u}$ and static pressure $p$ 
that are $x$-periodic. As we shall see, with these assumptions alone we may derive a version of the Kelvin 
minimum energy theorem. However, to derive the JA-relation and to guarantee that the potential flow velocity 
$\mathbf{u}_\phi$ is $C^\infty$ on the torus we must require further that the flow domain is reflection-symmetric about the 
spanwise-wall normal midplane at $x=0$, as shown in Fig.~\ref{scheme}\footnote{The reflection symmetry of 
the domain is unnecessary to derive our new JA-relation, if the flow is forced by prescribing constant values of 
$h_{in}$ on $S_{in}$ and $h_{out}$ on $S_{out},$ and these surfaces in that case may even be curved. This 
driving corresponds to solving the Poisson equation for pressure $p$ with mixed Dirichlet-Neumann conditions
of the form $p=h_{in}-(\frac{1}{2}|\mathbf{u}|^2+Q)$ on $S_{in},$ $p=h_{out}-(\frac{1}{2}|\mathbf{u}|^2+Q)$ 
on $S_{out},$ and $\partial p/\partial n=\nu \mathbf{n}\bm{\cdot}\Delta\mathbf{u}$ on $S_w.$ With the same 
boundary conditions for the new reference potential $\phi$ as in \eqref{mixedbc}, it is then easy 
to check that our proof of the Kelvin theorem and the JA-relation go through unchanged. However, in this alternative
construction there is no guarantee that $\mathbf{u}_\phi$ is $C^\infty$ at $S_{in}=S_{out}$}. Additionally, at the intersection 
of the sidewalls $S_w$ with inflow surface $S_{in}$ and outflow surface $S_{out}$ we assume that the normal 
vectors satisfy the geometric conditions 
\begin{align}
    \mathbf{n}_w\mathbf{\cdot n}_{in}=\mathbf{n}_w\mathbf{\cdot n}_{out}=0.
\end{align}
required for compatibility between Neumann conditions on $S_w$ and Dirichlet conditions on $S_{in}$ and $S_{out}.$ 


{ \begin{figure}
 \centering
\includegraphics[width=.48\textwidth]{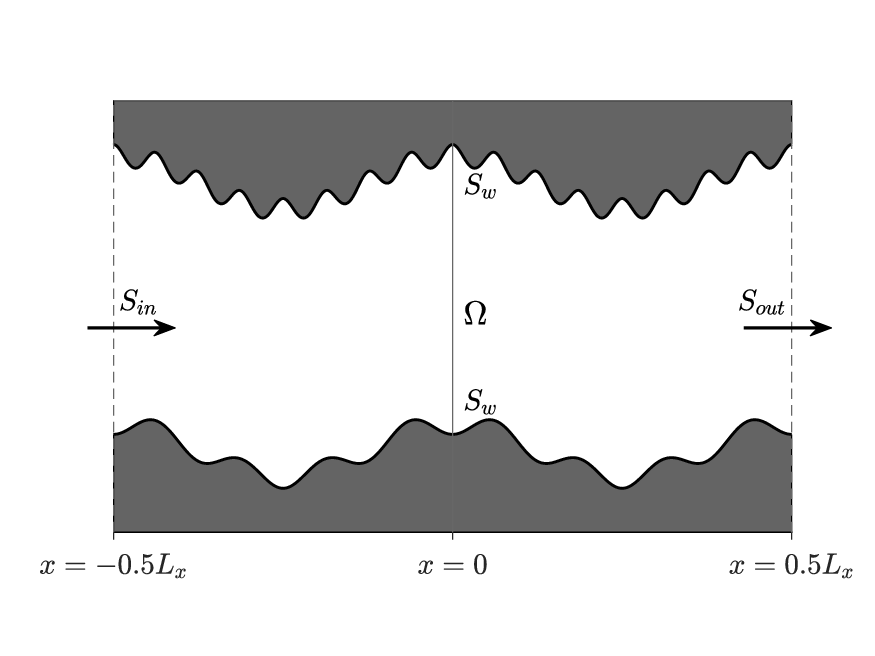}
\caption{Flow through a channel $\Omega$ with inflow surface $S_{in}$, outflow surface $S_{out}$, and sidewalls $S_w$. 
The domain is symmetric about the spanwise-wall normal plane at x=0.   }
\label{scheme}
\end{figure}}

Under these assumptions we define the potential $\phi$ of the reference Euler 
solution to satisfy the Laplace equation, $\nabla^2 \phi=0$, with  mixed 
Neumann-Dirichlet boundary conditions 
 \begin{align}\label{mixedbc} 
    \left.\frac{\partial \phi}{\partial n}\right|_{S_w}=0, \qquad 
    \phi\Big|_{S_{in}}=-\frac{1}{2}\Phi(t), \qquad
    \phi\Big|_{S_{out}}=+\frac{1}{2}\Phi(t).
\end{align}
For channel flow, we assume spanwise periodic boundary conditions as well. 
Here, the potential difference $\Phi(t)$ is chosen so that the Euler flow carries the entire mass flux, that is,
\begin{align}
    J_{\phi}(t):=\rho\int_{S_{in}}\mathbf{u_{\phi}}\cdot \hat{\mathbf{n}}\, dA=\rho\int_{S_{out}}\mathbf{u_{\phi}}\cdot \mathbf{n}\, dA\\
    =\rho\int_{S_{in}}\mathbf{u}\cdot \hat{\mathbf{n}}\, dA=\rho\int_{S_{out}}\mathbf{u}\cdot \mathbf{n} \, dA :=J(t).
\label{allJ} \end{align}
This potential is easily calculated by exploiting the homogeneity of the problem and first solving for $\phi_*=\phi/\Phi$, 
which satisfies the Laplace equation, $\nabla^2 \phi_*=0$ with mixed boundary conditions 
\begin{align}\label{potstar}
    \left.\frac{\partial \phi_*}{\partial n}\right|_{S_w}=0, \qquad
    \phi_*\Big|_{S_{in}}=-\frac{1}{2}, \qquad
    \phi_*\Big|_{S_{out}}=+\frac{1}{2}.
\end{align}
for which a unique solution exists. Now, let $\mathbf{u}_{\phi*}=\bm{\nabla}\phi_*$ and $J_*=\rho\int_{S_{in}}\mathbf{u}_{\phi*}\cdot \mathbf{n}\,dA$, leading to $J(t)=J_*\Phi(t)$ and  
\begin{align}
\phi= \Phi(t)\phi_*=J(t)\phi_*/J_*.
\end{align}
Observe that $\phi_*$ and consequently $J_*$ depend only on the channel geometry. The potential $\phi(t)$
that results for given $J(t)$ is uniquely defined, although in general it may depend upon the arbitrary 
choice of the surface $S_{in}=S_{out}$ in the periodic domain $\Omega$.  We remark in passing also that 
$E_\phi=(1/2)J\,\Phi,$ as shown by using as curvilinear coordinates the potential $\phi$ itself and any 
convenient parameterization of $\phi$-isosurfaces and by noting that $d\phi=|\mathbf{u}_\phi|d\ell$
for arclength $\ell$ along streamlines. 

The above construction yields a reference Euler solution $\mathbf{u_{\phi}}$ which is $C^\infty$
and $x$-periodic, when reflection symmetry of $\Omega$ about its midplane is assumed. 
Since $\bar{\phi}(x,y,z):=-\phi(-x,y,z)$ is another solution of the mixed boundary-value problem 
\eqref{mixedbc}, uniqueness of that solution implies the symmetry property
\begin{equation} \phi(-x,y,z)=-\phi(x,y,z). \label{phi-sym} \end{equation} 
This fact will be exploited together with the fact that the harmonic function $\phi\in C^\infty(\mathring{\Omega})$
where $\mathring{\Omega}=\Omega\backslash (S_{in}\cap S_{ou}\cap S_w)$ is the interior of the domain.
See \textcite{evans2010partial}, Theorem 2.6, p.28. Note that $\phi$ will in general be smooth up to $S_w$ if the 
sidewall $S_w$ is smooth, but we must show that all derivatives approaching $S_{in}=S_{out}$ from both sides agree. It is an elementary consequence of \eqref{phi-sym} that 
\be  \partial_x^m\phi(-\frac{L_x}{2},y,z)=\partial_x^m\phi(+\frac{L_x}{2},y,z), \quad\mbox{for all odd $m$}. \lb{oddm} \ee
We next show by induction that 
\be \partial_x^m\phi(-\frac{L_x}{2},y,z)=\partial_x^m\phi(+\frac{L_x}{2},y,z)=0, \quad\mbox{for all even $m\geq 2$}. \lb{evenm}\ee
For $m=2$ this follows by using the fact that $\phi$ is harmonic and is a spatial constant $\phi=\pm \Phi/2$
for $x=\pm L_x/2,$ so that  
$$ \partial_x^2\phi(\pm\frac{L_x}{2},y,z)=-(\partial_y^2+\partial_z^2)\phi(\pm\frac{L_x}{2},y,z)=0.  $$
We now assume that \eqref{evenm} holds for all even integers up to $m$ and then note 
that 
$$ \partial_x^{m+2}\phi(\pm\frac{L_x}{2},y,z)=-\partial_x^m(\partial_y^2+\partial_z^2)\phi(\pm\frac{L_x}{2},y,z)
=0 $$
by using the Laplace equation and the induction hypothesis, thereby completing the induction. It follows from
\eqref{oddm},\eqref{evenm} that $\mathbf{u}_\phi=\bm{\nabla}\phi$ is both $x$-periodic and $C^\infty$ in $\Omega.$

Note that the velocity potential $\phi$ itself obviously cannot be periodic, because of the anti-periodic b.c. 
\eqref{mixedbc}. On the other hand, it is not hard to show that $\phi_{per}:=\phi-\Phi x/L_x$ is 
$x$-periodic. In fact, this function solves the Laplace equation $\nabla^2\phi_{per}=0$ in $\Omega$
with the mixed boundary conditions
$$  \left.\frac{\partial \phi_{per}}{\partial n}=-\frac{\Phi}{L_x}n_x\right|_{S_w}, \qquad
    \phi_{per}=0\Big|_{S_{in}}, \qquad  
    \phi_{per}=0\Big|_{S_{out}}.  $$
The solution of this problem is unique and $x$-periodic, vanishing on $S_{in}=S_{out}.$ 
Furthermore, $\bm{\nabla}\phi_{per}=\bm{\nabla}\phi-\Phi/L_x$ so that the preceding discussion 
shows that $\phi_{per}$ is a $C^\infty$ function on the entire flow domain $\Omega.$ We thus conclude 
that $\phi$ is the sum of a (smooth) periodic function and a function linear in $x$:
$$\phi=\phi_{per}+\Phi x/L_x $$ 
This fact will prove important in our derivation below. 

With these results in hand we can essentially repeat the construction of \textcite{Huggins1970a}.
We note here just the key differences. One can define $\mathbf{u}_\omega=\mathbf{u}-\mathbf{u}_\phi$
as before, but now $\mathbf{u}_\omega$ satisfies the non-flow-through constraints \eqref{noflow-om} only at $S_w$ 
and not at $S_{in}=S_{out}$. However, the condition \eqref{allJ} that $\mathbf{u}_\phi$ carries the total mass flux 
still yields the weaker result that 
\begin{align}
\int_{S_{in}}\mathbf{u_{\omega}}\cdot\mathbf{\hat{n}}\,dA=\int_{S_{out}}\mathbf{u_{\omega}}\cdot\mathbf{n}\,dA=0.
    \lb{weak-cond} 
\end{align}
This suffices to imply that the potential and vortical fields are orthogonal, as the following brief calculation shows: 
\begin{align}
    & \int_{\Omega}\mathbf{u_{\phi}}\cdot\mathbf{u_{\omega}}\,dV=
    -\int_{S_{in}}\phi\mathbf{u_{\omega}}\cdot \mathbf{n} \,dA
    -\int_{S_{out}}\phi\mathbf{u_{\omega}}\cdot \mathbf{n} \,dA \nonumber \\
    & \hspace{90pt} -\int_{S_w}\phi\mathbf{u_{\omega}}\cdot \mathbf{n} \,dA \nonumber \\ 
    &\hspace{20pt} =\frac{1}{2}\Phi(t)\int_{S_{in}}\mathbf{u_{\omega}}\cdot \mathbf{n} \,dA
    -\frac{1}{2}\Phi(t)\int_{S_{out}}\mathbf{u_{\omega}}\cdot \mathbf{n} \, dA=0.
\end{align}
Note that neither smoothness of $\phi$ nor even flatness of the sections $S_{in},$ $S_{out}$ were required here. 
The other key step in the derivation of \textcite{Huggins1970a} where the constraints \eqref{noflow-om} were 
used was in the calculation \eqref{e_omega} of the balance equation for $E_\omega,$ where they 
were invoked to eliminate the boundary terms at $S_{in}$ and $S_{out}$ involving $h_\omega.$ 
In fact, the weaker conditions \eqref{weak-cond} again suffice, if one recalls that 
\be h_\omega=h+\partial_t\phi=p +\frac{1}{2}|\mathbf{u}|^2+Q+\partial_t\phi \lb{hom2} \ee
so that $h_\omega$ is the sum of a smooth, $x$-periodic part $h_{\omega,per}=
p +\frac{1}{2}|\mathbf{u}|^2+Q_{per}+\partial_t\phi_{per}$ and a linear part 
$h_{\omega,lin}=\dot{\Phi}x/L_x-\gamma x.$ In that case, the periodic part gives 
no contribution and the linear part contributes zero also because 
\bea && \rho\int_{S_{in}}h_{\omega,lin}\mathbf{u}_{\omega}\cdot\mathbf{n}\,dA
        +\rho\int_{S_{out}}h_{\omega,lin}\mathbf{u}_{\omega}\cdot\mathbf{n}\,dA  \cr
       &&\hspace{10pt} = -\frac{1}{2} (\dot{\Phi} -\gamma L_x) \rho\int_{S_{in}} \mathbf{u}_{\omega}\cdot\mathbf{n}\,dA \cr
       &&\hspace{30pt}  +\frac{1}{2} (\dot{\Phi} -\gamma L_x) \rho\int_{S_{out}}\mathbf{u}_{\omega}\cdot\mathbf{n}\,dA  
       \ = \ 0 .      
        \eea
Here we required flatness of $S_{in},$ $S_{out}$ so that $h_{\omega,lin}$ is constant on those surfaces and
continuity of $\mathbf{u}_\omega=\mathbf{u}-\mathbf{u}_\phi$ at $S_{in}=S_{out}$ 
to cancel the contribution from $h_{\omega,per}.$
In conclusion, the balance equation \eqref{e_omega} for $E_\omega$ again holds, and all of the rest of the derivation
is identical to that of \textcite{Huggins1970a}.       

There are a few further simplifications compared with the construction of \textcite{Huggins1970a}
due to the fact that both $h$ and $h_\omega$ are now smooth, $x$-periodic functions plus a part which is 
linear in $x.$ Thus, rate of work $\mathcal{W}$ by total pressure head defined in \eqref{W-def} now becomes
$$ \mathcal{W}= \gamma(t) L_x J= (\Delta h) J $$
where we have defined $\Delta h=\gamma L_x$ as the drop in total pressure. Likewise, the work done 
by the total rotational pressure is 
$$ \mathcal{W}_\omega= [(\Delta h) - \dot{\Phi}] J = (\Delta h_\omega) J, $$
so that the detailed $JA$-relation now becomes simply 
\begin{align}\label{detailedJA}
\mathcal{T}=&(\Delta h_\omega) J=(\Delta h) J-\mathcal{W}_\phi,
\end{align}
The rate of work by the potential flow simplifies also as
\begin{align}
\mathcal{W}_\phi=& J\dot{\Phi} = J\dot{J}/J_* = \dot{E}_\phi,
\end{align}
For the special case of a flow with a mass flux constant in time, $dJ/dt=0$, one gets 
furthermore $\Delta h_{\omega}=\Delta h$ and $\mathcal{T}=(\Delta h) J$.

It is also instructive to consider the canonical case of channel flow with flat plane-parallel walls
and $\mathbf{f}=\mathbf{0}$. In that case, as previously noted, our construction yields $\phi(t)=U(t)x$ and 
$\mathbf{u}_\phi=U(t)\hat{{\bf x}}$ is spatially 
constant. It follows then from the alternative formula \eqref{Tdef2} for $\mathcal{T}$ in the 
Introduction that
\be  \mathcal{T}= \rho U \int_{S_w} \tau_{xy}^w\, dA, \lb{Ttau} \ee
which is the energy dissipated by viscous wall drag. Thus, in this particular case, both the work done 
against rotational pressure and the dissipation by drag are instantaneously related to vorticity flux 
across the channel. The transfer term likewise simplifies to
$$ \mathcal{T} = -\rho U \int_\Omega \Sigma_{yz} \, dV 
= -\rho U\int_\Omega (\omega_z v-\omega_y w - \nu \partial_y\omega_z) \, dV $$ 
where note that $\int_\Omega \partial_z\omega_y \, dV =0$ because of the spanwise 
periodic b.c. Note further because of the vector calculus identity 
\be \omega_z v-\omega_y w =-\partial_x(u^2)-\partial_y(vu)-\partial_z(wu)
     +\frac{1}{2}\partial_x(u^2+v^2+w^2)\ee 
and the assumed boundary conditions that the net contribution to $\mathcal{T}$ from the nonlinear
term {\it vanishes}, when integrated over the flow volume. This vanishing value is special to channel flow 
with flat, parallel walls, because of the high degree of symmetry of this flow, whereas the nonlinear 
contribution to the JA-relation is generally not zero (e.g. see next section). Integrating the remaining term 
$\nu \partial_y\omega_z$ in $y$
directly recovers \eqref{Ttau} and the detailed JA-relation reduces 
to an instantaneous version of the time-average result for turbulent channel flow, $\langle \Sigma_{yz}\rangle =-u_\tau^2/h,$ 
previously discussed in the literature \cite{huggins1994vortex,eyink2008,kumar2023flux}. Note 
however that the time-average $\langle\omega_z v-\omega_y w \rangle(y)\neq 0$ and this term is crucial 
to give a $y$-independent constant mean total flux, although contributions from negative and positive signs 
of the mean nonlinear flux exactly cancel when integrated over $y$-locations. See \textcite{kumar2023flux} for 
more discussion of the physical mechanisms.\\

\section{Numerical Results for a Flat-Wall Channel with A Smooth Bump}

In this section we present a numerical application of our new detailed JA relation. In order 
to investigate flow separation and its contribution to drag, we have selected for study a streamwise-periodic 
channel flow with plane-parallel walls modifed by addition of a smooth bump or ridge at the wall,  
with a cosine profile in the streamwise direction over a complete period, from minimum to minimum, 
and spanwise constant. See Fig.~\ref{phistar} for a sideview of the geometry. 
We keep the bulk flow velocity $U$ constant, for ease of demonstration, with the 
pressure gradient $\gamma(t)$ which drives the flow varying to maintain the constant flow rate. We take the $x$-direction
as streamwise, $y$-direction as wall-normal and and $z$-direction as spanwise. We consider a domain of 
size $(L_x,L_y,L_z)=(1,1,0.5)$ in arbitrary units  and the height and width of the cosine bump are $0.1L_x$ 
and $0.5L_x$, respectively. We initialize the velocity field with a constant value $\mathbf{u}=(U,0,0)$,
The Reynolds number based on bulk velocity $U$ and channel height $L_y$ is thus constant at 
$Re=UL_y/\nu =975$, which results in an 
unsteady laminar flow, sufficient to drive flow separation from the bump
and to generate a rotational wake. 

\begin{table}[h]
    \centering
    \begin{tabular}{l c c c c} 
        \toprule
        Case & Nx & Ny & Nz & $RelErr(\%)$ \\
        \midrule
        1 & 50  & 50  & 25  & 11.7 \\
        2 & 80  & 80  & 40  & 7.88 \\
        3 & 160 & 160 & 81  & 4.38 \\
        4 & 216 & 216 & 108 & 3.56 \\
        \bottomrule
     \end{tabular}
            \caption{The effect of grid size on maximum error} \label{tab:example_table}
\end{table}

To compute this flow numerically we use the laminar pimpleFoam~\cite{penttinen2011} solver from 
OpenFOAM~\cite{welleretal1998}, with a body-fitted structured mesh of hexahedral cells and a range of 
mesh sizes listed in Table~\ref{tab:example_table}. A convergence study shows that the results are accurate 
within a few percent for the finest mesh $(N_x,N_y,N_z)=(216,216,108)$ (see below) and all concrete 
results presented here are for that resolution. Numerical field values are output at time intervals 
of $\Delta tU/L_x =0.195 $ starting at $tU/L_x=0.195.$ 
Our goal is 
to numerically evaluate the detailed JA-relation \eqref{detailedJA}, which here takes the concrete form 
\be \gamma(t) L_x J= -\int_{\Omega} \rho\mathbf{u}_{\phi}\cdot (\mathbf{u}\times \bm{\omega}-\nu \bm{\nabla\times}\bm{\omega})\, d V :=\mathcal{T}(t) \label{JAcomp} \ee
since $dJ/dt=0$ and $\mathbf{f}=\mathbf{0}.$ The volume-integral in \eqref{JAcomp} was computed numerically 
by a Riemann sum where each cell is associated with a single value and all values are multiplied by cell volume and added to get integrals. To obtain $\mathbf{u}_\phi=\Phi(t)\mathbf{u}_\phi^*,$  the geometry-dependent dimensionless potential 
$\phi_*$ satisfying b.c. \eqref{potstar} was calculated by solving the Laplace equation using the same mesh. 
The results are shown in Fig.~\ref{phistar}, which plots $\phi_*$ as a color map and representative streamlines.
The prefactor $\Phi(t)$ is time-independent for this flow with constant bulk velocity and fixed by the relation 
$\Phi=J/J_*.$ All space-gradients such as $\bm{\omega}=\bm{\nabla\times}\mathbf{u}$ and $\bm{\nabla\times\omega}$
were calculated by central differences. We find that the maximum relative error between the LHS and RHS of the JA relation 
\eqref{JAcomp}, or $RelErr(\%)  = 100\max_t\left|1-\frac{\mathcal{T}(t)}{JL_x\gamma(t)}\right|$ decreases with 
mesh resolution, as shown in the final column of Table~\ref{tab:example_table}. We deemed the maximum error 
$\leq 3.56\%$ achieved at our highest resolution to be adequate for the purposes of this study.  

More detailed information about accuracy is afforded by the plots in Fig~\ref{ts} of the time series of the driving 
pressure-gradient $\gamma(t)$ and of the transfer term $\mathcal{T}(t)$ in the detailed JA-relation, suitably 
non-dimensionalized, which agree quite well over the entire recorded time period. However, in addition to numerical 
validation, further information about the physics is provided by the plots in Fig~\ref{ts} of the separate 
contributions to $\mathcal{T}(t)$ arising from viscous and nonlinear vorticity transport. At the moderate 
Reynolds number of the simulation, the viscous contribution is largest and the nonlinear contribution only 
about half as large. On the other hand, the instantaneous drag as measured by $\gamma(t)$ exhibits 
distinctive oscillations, which are contributed entirely by the nonlinear transport term  in $\mathcal{T}(t)$
whereas the viscous term decays monotonically in time. We argue that the local maxima in drag are due to periodic 
episodes of strong vortex shedding from the smooth bump, whereas the local minima are due to episodes of 
weaker shedding. We present several pieces of evidence to support this interpretation.

{ \begin{figure}[h!] 
 \centering
\includegraphics[width=0.4\textwidth]{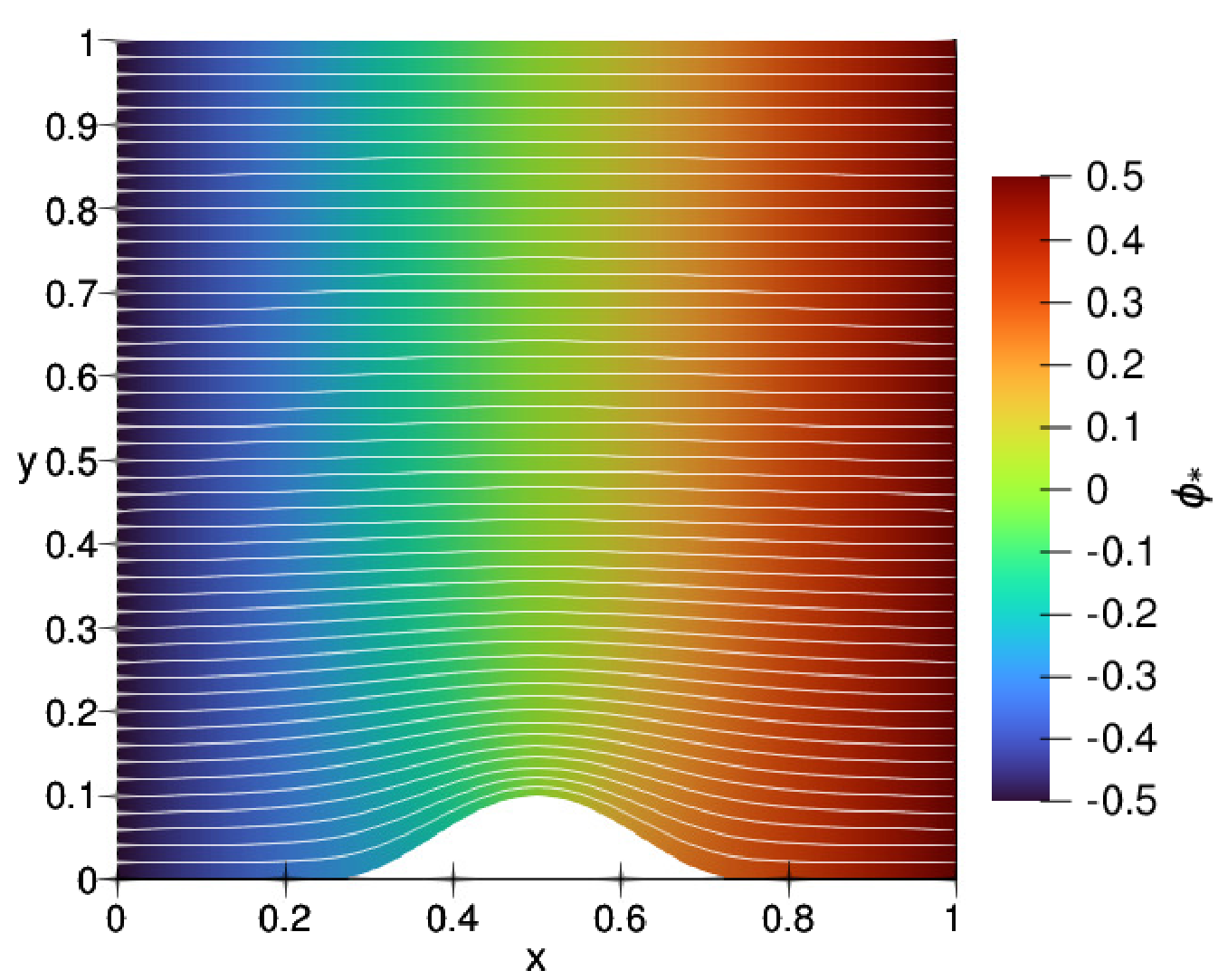}
\caption{Dimensionless background potential $\phi_*$ and its streamlines.}
\label{phistar}
\end{figure}

\begin{figure}[h!]
 \centering
\includegraphics[width=0.45\textwidth]{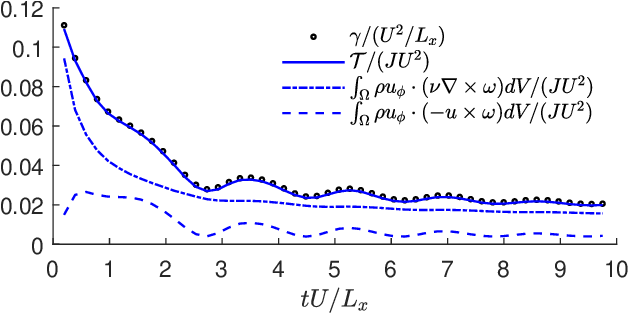}
\caption{Time series of terms in the detailed JA relation (Eq.~\ref{JAcomp}) for flow past a bump in a periodic channel with a constant flow rate ($dJ/dt=0$), also showing separate viscous and nonlinear contributions to the transfer term.}
\label{ts}
\end{figure}

\begin{figure}[h!]
 \centering
\includegraphics[width=0.45\textwidth]{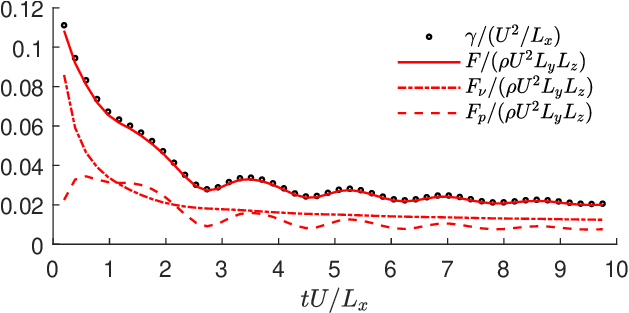}
\caption{Time series of terms in the global momentum balance, the net drag force $F_x(t)$ on the wall and the instantaneous pressure gradient $\gamma(t)$, also showing separate contributions from skin friction and pressure to drag force.}
\label{forces}
\end{figure}} 

One such piece of evidence comes from an additional exact relation derived from the constraint 
$\frac{1}{L_xL_yL_z}\int_\Omega\mathbf{u}\,dV=U\hat{\mathbf{x}}$ and the global 
momentum balance obtained by integrating the governing Navier-Stokes equation over the flow domain: 
\be \int_{S_w} (-P\mathbf{n}+\rho \bm{\tau}_w)\, dA + \dot{J}(t)L_x
=\rho\gamma(t) L_x L_yL_z \hat{\mathbf{x}}. \label{mombal0} \ee 
In the case considered here $dJ/dt=0,$ so that global momentum balance reduces to the relation 
\be \mathbf{F}(t):=\int_{S_w} (-P\mathbf{n}+\rho \bm{\tau}_w)\, dA
=\rho\gamma(t) L_x L_yL_z \hat{\mathbf{x}}, \label{mombal} \ee 
where the lefthand side is the instantaneous drag force exerted by the fluid on the channel walls and the righthand
side is the instantaneous force applied by the external pressure gradient to the fluid. 
Plotted in Fig.~\ref{forces} are the times series of the $x$-components of the two sides of Eq.\eqref{mombal}, 
suitably normalized, whose excellent agreement again validates our numerical solution. More physically informative are the plots in Fig.~\ref{forces} of the separate contributions to the drag force from the viscous skin friction and the 
pressure (form drag), which show remarkably similar (but not identical) behaviors as the viscous and nonlinear 
transport contributions to the JA-relation as plotted in Fig.~\ref{ts}. The similarity of the viscous contributions 
is unsurprising, as we have already noted in Eq.\eqref{Tdef2} that 
\be 
\int_{\Omega} \mathbf{u}_{\phi}\cdot \nu \bm{\nabla\times}\bm{\omega}\, d V = 
\int_{\partial \Omega} \mathbf{u_{\phi}}\cdot \bm{\tau}_w \, dA.  
\lb{parts} \ee 
by a simple application of the divergence theorem. Thus, the viscous term in the JA transfer 
term coincides with the viscous term in the drag force after substituting $\mathbf{u}_\phi$ for 
$U\hat{\mathbf{x}}.$ Since Fig.~\ref{phistar} shows that $\mathbf{u}_\phi$ and $U\hat{\mathbf{x}}$
are quite similar, it is understandable that the two viscous contributions are closely correlated.

\begin{figure}[h!]
      \centering
      \begin{subfigure}[b]{0.32\textwidth}
         \centering
         \includegraphics[width=\textwidth]{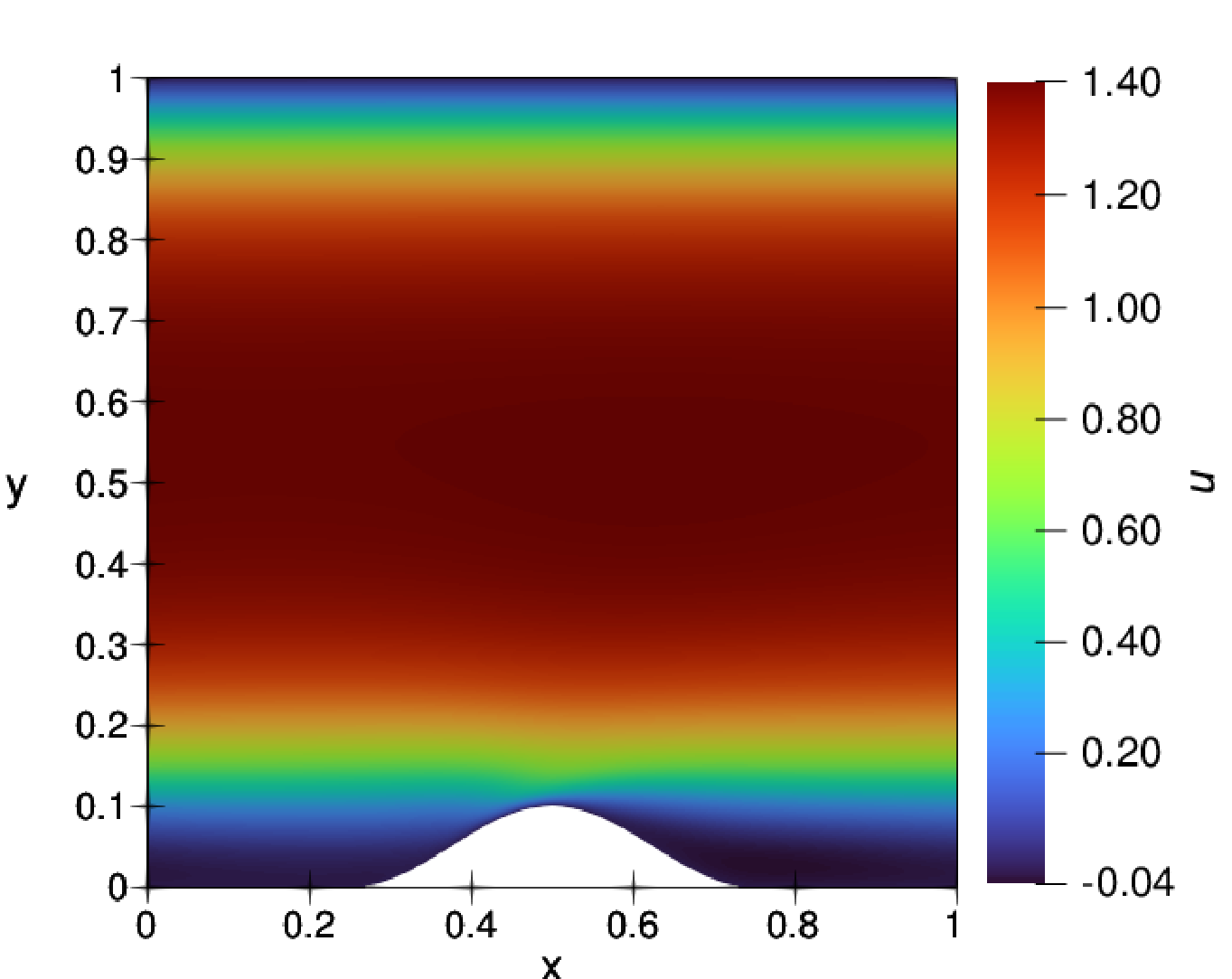}
         \caption{Streamwise velocity }
         \label{uinst}
         \end{subfigure}
         \vfill
     \begin{subfigure}[b]{0.32\textwidth}
         \centering
         \includegraphics[width=\textwidth]{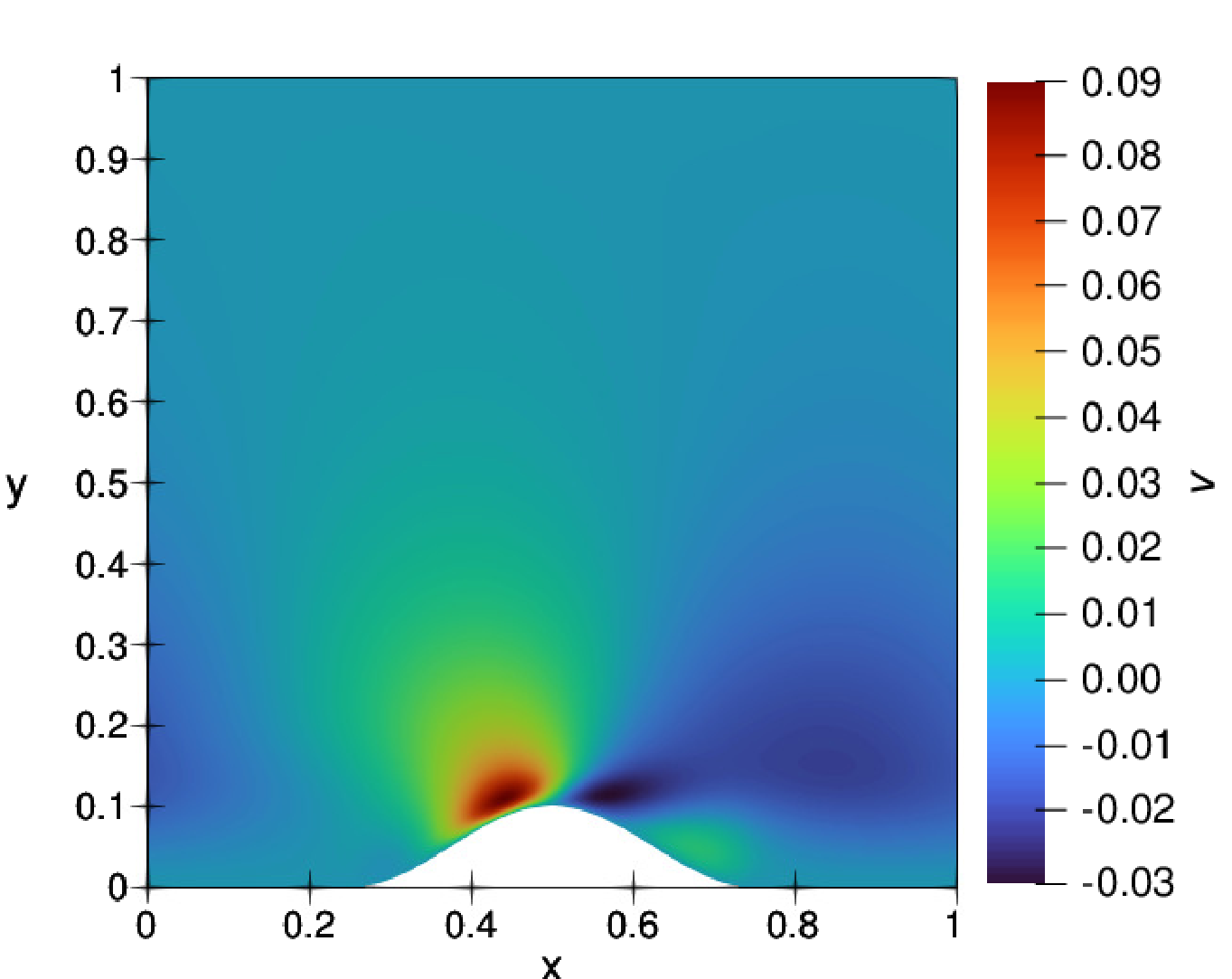}
        \caption{Wall-normal velocity }
         \label{vinst}
     \end{subfigure}
        \vfill
        \begin{subfigure}[b]{0.32\textwidth}
        \centering
        \includegraphics[width=\textwidth]{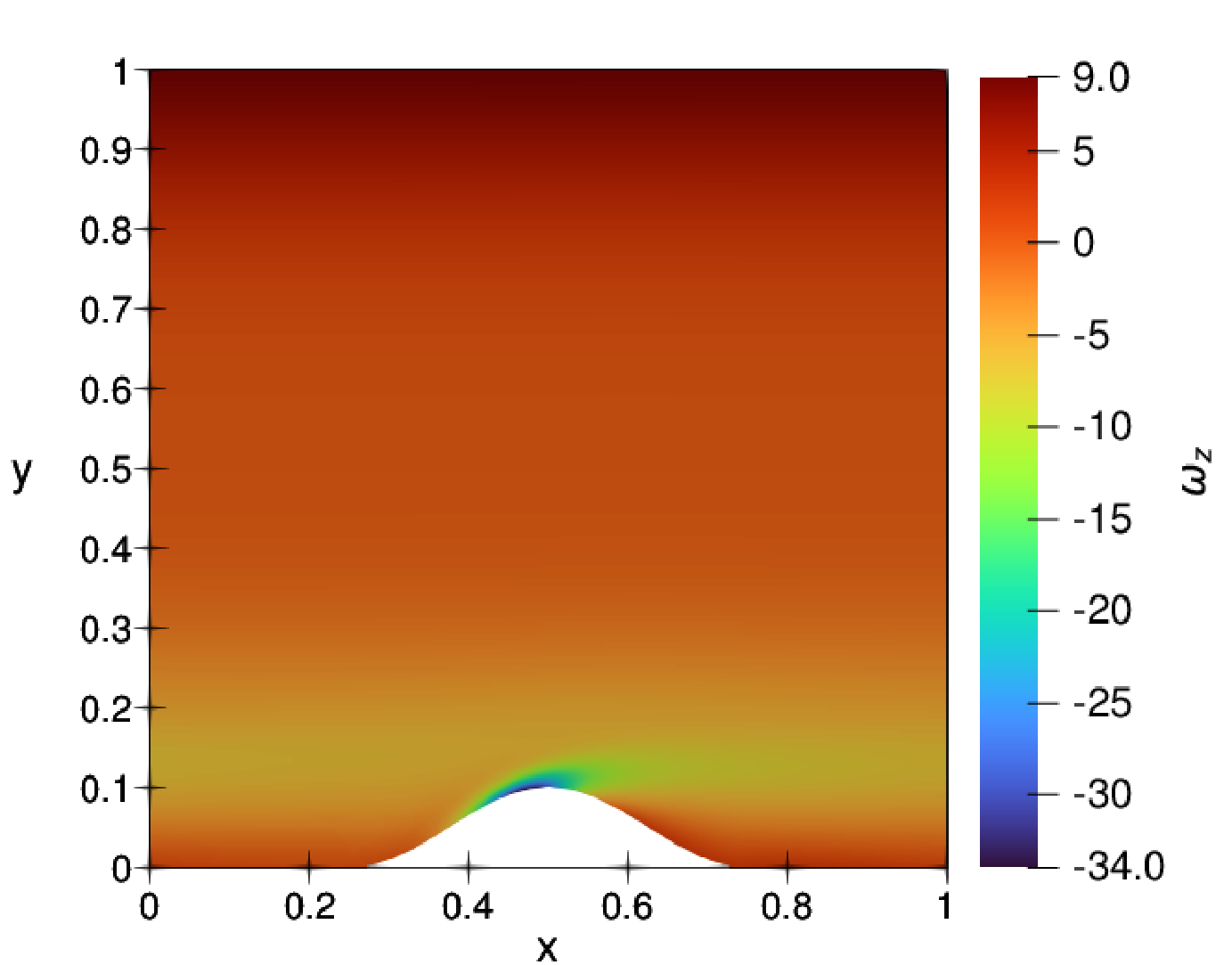}\caption{Spanwise vorticity }
        \label{ozinst} 
     \end{subfigure}
              \caption{ Instantaneous velocity fields, (a) streamwise and (b) wall-normal, normalized by $U$, 
              and (c) spanwise vorticity field normalized by $U/L_x$, at $tU/L_x=9.75$, $z=0.25$. }   \label{fields}
\end{figure}

We cannot find any such direct correspondence between the nonlinear term $\mathcal{T}_{nlin}(t)$ and 
the form drag $F_{px}(t)$, but it is well-known that large form drag is associated to earlier or stronger 
shedding of vorticity by flow separation. Thus, the similar oscillations observed in both the form drag
and the nonlinear transfer term are likely both due to oscillations in separation.  Boundary-layer separation 
can in fact be verified in this flow by visualization of spatial fields in Fig~\ref{fields}. For simplicity
we have chosen to visualize a late time $tU/L_x=9.75$ when the flow has become nearly steady and we plot fields in the vertical
$xy$-plane at the spanwise midsection $z=0.25$. The plot of the streamwise velocity $u$ in Fig.~\ref{uinst}
is relatively uninformative, showing just a slightly elavated region of reduced streamwise velocity downstream 
of the bump. However, the plot of the wall-normal velocity $v$ in Fig.~\ref{vinst} shows a clear upward jet 
just upstream of the bump, while just downstream there is a bipolar pattern of downflow followed by upflow 
indicative of a recirculation bubble. Most compelling is the plot of the spanwise vorticity in Fig.~\ref{ozinst}
which shows a strong sheet of negative spanwise vorticity on the upstream face of the bump associated to a 
viscous boundary layer which is then shed into the flow downstream of the bump. On the downstream face of the 
bump the vorticity is instead positive, indicating a recirculation bubble. In fact, we see such 
clear evidence of flow separation at all recorded times. 

\begin{figure}[h!]
      \centering
      \begin{subfigure}[b]{0.32\textwidth}
         \centering
         \includegraphics[width=\textwidth]{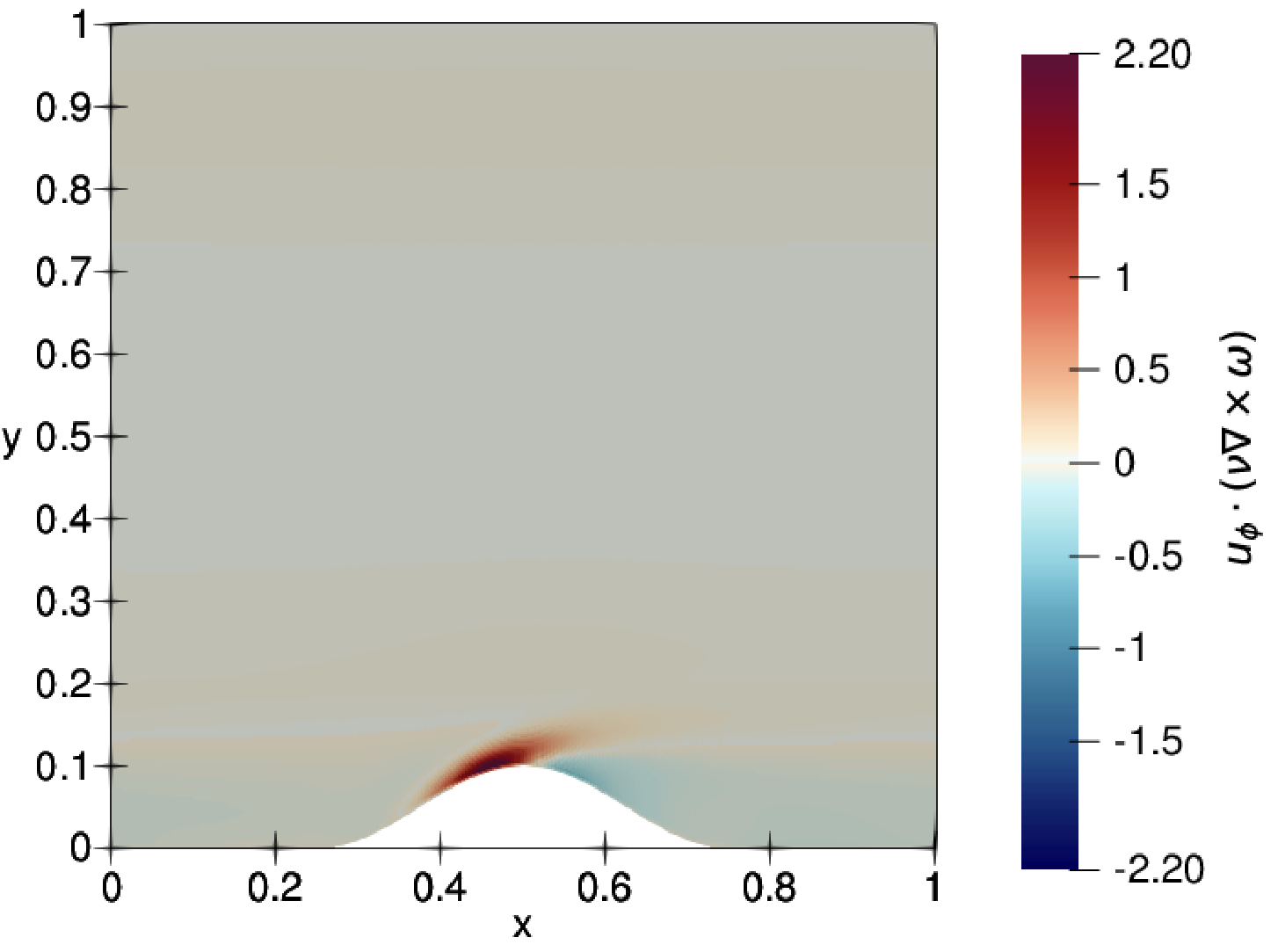}
        \caption{Viscous contribution to $\mathcal{T}$-integrand, accounting for $78.6\%$ of instantaneous drag.}
         \label{visc}
     \end{subfigure}
     \hfill
         \begin{subfigure}[b]{0.32\textwidth}
         \centering
         \includegraphics[width=\textwidth]{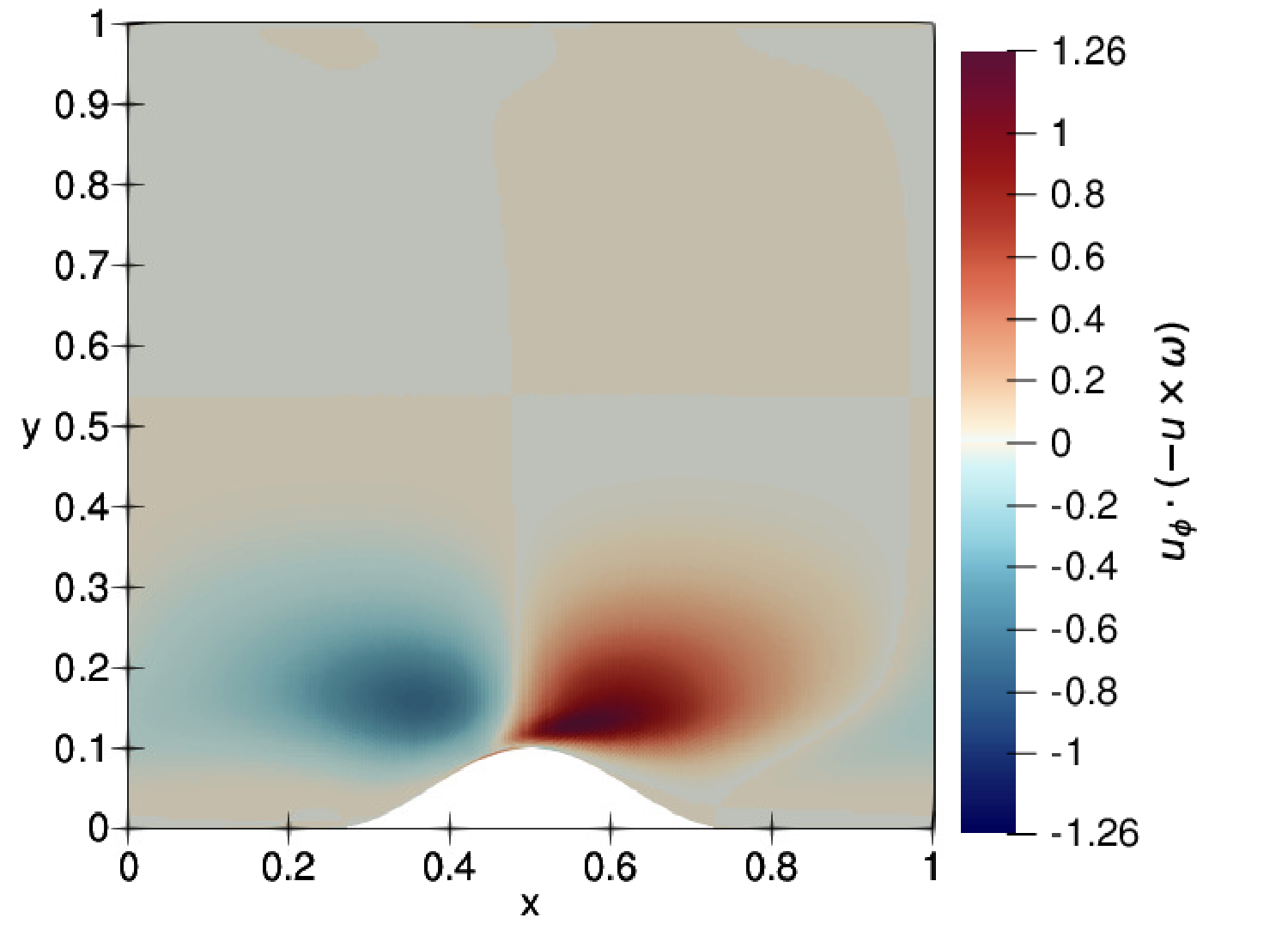}\caption{Nonlinear contribution to $\mathcal{T}$-integrand, 
          accounting for $21.4\%$ of instantaneous drag. }
         \label{nl}
     \end{subfigure}
     \hfill
     \begin{subfigure}[b]{0.32\textwidth}
         \centering
         \includegraphics[width=\textwidth]{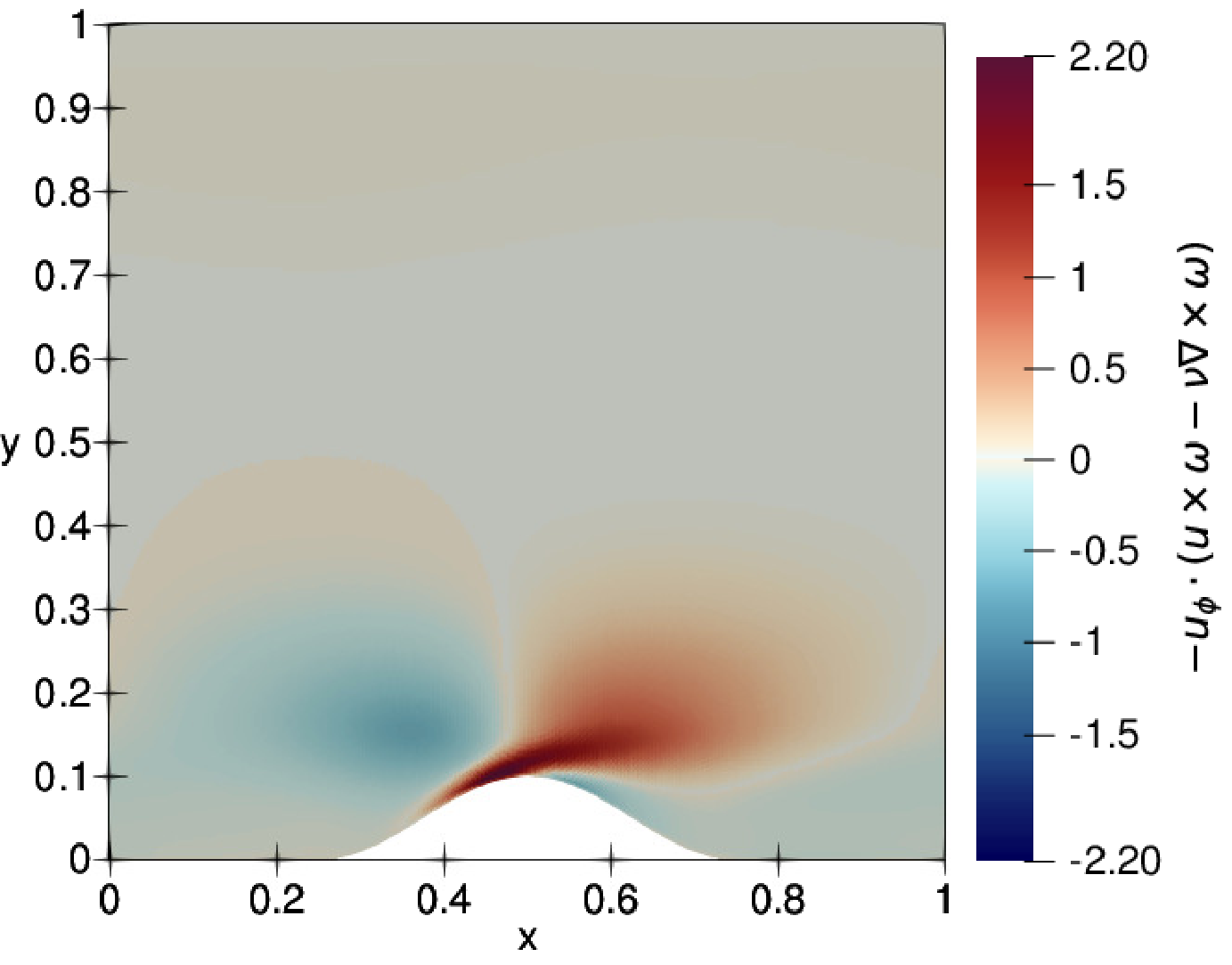}
         \caption{$\mathcal{T}$-integrand in Eq.~\eqref{JAcomp}, representing local flux of spanwise vorticity across potential flowlines. }
         \label{syz}
    \end{subfigure}
                 \caption{Instantaneous fields of (c) the integrand of $\mathcal{T}$ in the detailed JA-relation \eqref{JAcomp}, and (a) viscous and (b) nonlinear contributions to the integrand, all normalized by $\rho U^3/L_x$. Fields are shown for 
                 $tU/L_x=9.75,$ $z=0.25$. }   \label{fluxcontri}
\end{figure}

To get physical understanding of the relation of drag to such vorticity dynamics, we can visualize the integrand 
appearing in the spatial integral which defines the transfer term $\mathcal{T}(t)$ in the detailed JA-relation 
of Eq.\eqref{JAcomp}. We plot this integrand in Fig.~\ref{fluxcontri} at the same time $tU/L_x=9.75$ and in the 
$xy$-plane at the same spanwise position $z=0.25$ as the flow fields plotted in Fig.~\ref{fields}, so that 
the two may be compared directly. We note, however, that while our flow varies substantially in time, it is rather 
spanwise homogeneous, so that the plots in $xy$-planes at other spanwise positions are very similar. 
We plot in Fig.~\ref{visc} the viscous contribution to the integrand, in Fig.~\ref{nl} the nonlinear 
contribution, and in Fig.~\ref{syz} the combined integrand, representing local total flux of 
vorticity across flowlines of the Euler potential. We discuss each of the plots in turn. 

The viscous contribution to the JA-transfer term plotted in Fig~\ref{visc} can be readily understood,
because Huggins' flux tensor $\bm{\Sigma}$ appearing in \eqref{Tdef} for flux normal to the wall 
is exactly equal to the Lighthill boundary vorticity source $\bm{\sigma}$ , or 
\be \bm{\Sigma}^\top\mathbf{n}=\bm{\sigma}=\nu \mathbf{n}\times(\bm{\nabla \times\omega}), \label{lyman} \ee
where the relevant expression for $\bm{\sigma}$ is that of \textcite{lyman1990} rather than the alternative expression 
of \textcite{lighthill1963} and \textcite{panton1984incompressible}. Thus, the viscous vorticity 
flux in the flow interior directly continues that from the solid wall. Crucially, all transfer terms 
plotted in Fig.~\ref{fluxcontri} arise from {\it wall-normal flux of spanwise vorticity}, since the  potential 
flow-lines are parallel to the wall and furthermore $\bm{\omega\cdot}\mathbf{n}=0$ and 
$\mathbf{n}\bm{\cdot\sigma}=0,$ i.e. wall-normal vorticity and its fluxes are negligible in the vicinity 
of the surface. However, as also emphasized
by \textcite{lighthill1963} and especially by \textcite{morton1984}, vorticity generation at the surface
is an essentially inviscid process driven by tangential pressure gradients, as shown by the equivalent 
formula 
\be \bm{\sigma}=-\mathbf{n}\bm{\times\nabla}p. \label{lightmort} \ee 
Thus, the favorable pressure-gradient on the upstream side of the bump generates negative 
spanwise vorticity, whereas the adverse pressure-gradient on the downstream side generates positive 
spanwise vorticity. For plots of the pressure fields, see Supplementary Materials, \S IV. 
These signs are observed both in the plot of spanwise vorticity in Fig~\ref{ozinst}
and in the plot of the viscous transfer in Fig.~\ref{visc}. As emphasized by \textcite{lighthill1963}, 
however, the change of sign of $\sigma_z$ occurs earlier than the change of sign of $\omega_z$ (the point of separation), 
because it takes some time for the reversed positive flux to subtract the negative vorticity already present,
and this delay is clearly observed in Figs.~\ref{ozinst} \& \ref{visc}. The viscous transport 
of spanwise vorticity into the flow interior continues that at the surface but decreases 
rapidly as vorticity gradients drop off. 


The nonlinear contribution to the JA-transfer term plotted in Fig~\ref{nl} is the dominant one through 
the bulk of the flow, but consists of two large lobes of opposite sign upstream and downstream of the bump,
which substantially cancel. Thus, at the moderate Reynolds number of this simulation, nonlinear transfer provides 
only 21.4\% to the instantaneous drag at $tU/L_x=9.75$ and viscous transfer the remaining 78.6\%. The dominant 
contribution to the nonlinear vorticity flux in the region above the bump is the streamwise advection of 
spanwise vorticity, $\Sigma_{xz}\simeq u\omega_z,$ as may be seen from the plots in Fig.~\ref{fields}.  
The streamwise velocity plotted in Fig.~\ref{uinst} is more than an order of magnitude larger than the 
wall-normal component in Fig.~\ref{vinst}, while the spanwise velocity (not shown) is even smaller. The largest 
component of vorticity is by far the spanwise one $\omega_z$ and, in the region just above the bump, its sign is negative.
This is the dominant sign of vorticity shed from the bump which then, given the periodic boundary conditions, 
recirculates through the domain in the streamwise direction. Note, incidentally, that the dominant shedding 
of negative spanwise vorticity is directly related to form drag on the bump by the Lighthill-Morton relation 
\eqref{lightmort}, since the smaller flux of positive vorticity after separation implies that the pressure 
never fully recovers its upstream value. The flux $\Sigma_{xz}$ contributes to transfer 
across the potential streamlines because the latter bend vertically upward just upstream of the bump 
and vertically downward just downstream; see Fig.~\ref{phistar}. These considerations easily account 
for the observed signs of the two lobes in Fig~\ref{nl}\footnote{A quick way to check this is to observe that 
the corresponding wall-normal component of the Lamb vector, $(\mathbf{u}\bm{\times\omega})_y,$ points
upward, and it thus aligns with $u_{\phi y}>0$ upstream of the bump but anti-aligns with $u_{\phi y}<0$ downstream.}.  
The reason that the positive/drag-producing lobe downstream dominates over the negative/drag-reducing lobe upstream
is that the streamwise vorticity is strongest immediately after it is shed, whereas the vorticity 
periodically re-entering the flow domain upstream is diffused and weaker.

Combining the viscous and nonlinear contributions yields the total transfer integrand plotted 
in Fig~\ref{syz}. A very simple and intuitive picture thereby emerges for the origin of drag via 
vorticity dynamics. Negative spanwise vorticity is generated by the favorable pressure gradient 
on the upstream side of the bump, while a smaller amount of positive vorticity is generated 
by the adverse pressure gradient downstream. This vorticity viscously diffuses into the flow interior 
where nonlinear advection then takes over, convecting the excess negative spanwise vorticity downstream. 
Drag is produced as the negative spanwise vorticity crosses the streamlines of the background Euler potential. 
This picture directly relates the nonlinear flux contribution in the JA-relation to form drag,
since the latter results from the shedding of excess negative spanwise vorticity, and we can therefore 
understand the high correlation between the two terms observed in Figs.~\ref{ts} \& \ref{forces}. 
Note that the results that we have observed here for $tU/L_x=9.75$ are quite general and hold 
at all recorded times. Only the strength of vortex shedding varies with time, with strong shedding 
at times of local maximum drag in Figs.~\ref{ts} \& \ref{forces} and weak shedding at times of local 
minimum drag. See Supplemental Material, \S II-III, in particular for a comparison of the two times 
$tU/L_x=3.51$ and $tU/L_x=4.485$ corresponding to a local maximum and minimum, respectively. 

Although we have considered only a single flow geometry at a single Reynolds number, many 
of our conclusions are much more general. In fact, the JA-relation has recently been 
evaluated by \textcite{du2023vorticity} for external flow past spherical  and spheroidal bodies
and their results are very similar to ours. Spanwise (azimuthal) vorticity is generated in that flow  
principally by favorable pressure gradients on the body surface. This azimuthal vorticity 
diffuses outward from the sphere by viscosity but is shed rapidly into the flow  
by boundary-layer separation. Nonlinear advection takes over, with convection, stretching,
and twisting of vorticity, and resultant drag is produced by the integrated flux of the 
azimuthal vorticity across the streamlines of the background Euler potential. One difference is that 
\textcite{du2023vorticity} do not see an ``anti-drag'' lobe upstream of the body similar to ours in 
Fig.~\ref{syz}, because they do not use periodic boundary conditions and their inflow has negligible 
vorticity. In addition, their simulations are at higher Reynolds numbers than ours and the 
wake behind their body is fully turbulent. The vorticity dynamics in wall-bounded turbulent 
flows is more complex than what we observe in our laminar flow. As just one example, we observe
nonlinear vorticity transfer in our flow to be dominated by streamwise advection of spanwise vorticity
across potential streamlines, but the spanwise transport of wall-normal vorticity is found to play an 
essential role in turbulent channel-flow in the buffer layer and throughout the log-layer, related to 
velocity-correlated vortex-stretching\cite{kumar2023flux}. Nevertheless, the drag 
in turbulent wall-bounded flows is due also to the cross-stream flux of spanwise 
vorticity\cite{huggins1994vortex,eyink2008,kumar2023flux}. Thus, the Josephson-Anderson 
relation reveals a deep underlying unity in the origin of drag via vorticity dynamics, 
encompassing flows both internal and external, both laminar and turbulent, both 
classical and quantum. 

\section{Conclusions}

We have reviewed in this paper the detailed Josephson-Anderson relation for instantaneous drag
first derived by \textcite{Huggins1970a} for internal flows through general channels and we have 
explained how this result provides the exact analogue of the drag formulas for external flow past bodies 
derived by \textcite{wu1981theory}, \textcite{lighthill1986fundamentals,lighthill1986informal}, 
\textcite{howe1995force}, \textcite{eyink2021}, and others\cite{biesheuvel2006force}. In all of these works, 
instantaneous drag is divided into a potential part and an ``effective" rotational part that arises from vorticity 
flux across streamlines of the background potential Euler flow. However, we showed that the original 
relation of \textcite{Huggins1970a} suffers from significant problems when applied to classical turbulence 
and, in particular, his prescription for the background potential introduces a spurious vortex sheet for 
the streamwise periodic flows that are widely employed in numerical simulations. We proposed instead a 
reference potential Euler flow whose mass flux matches that of the total velocity field, while also ensuring 
that the vortical and potential velocity fields are orthogonal. The main theoretical result of our paper 
is the new detailed Josephson-Anderson relation \eqref{detailedJA} for streamwise periodic flows, which 
equates the instantaneous rate of work $\mathcal{W}_\omega$ due to rotational pressure, given by \eqref{Wom}, 
and the integrated flux of vorticity $\mathcal{T}$ across potential streamlines, given by \eqref{Tdef}.  
We finally illustrated the utility of this relation by the example of Poiseuille flow in a flat-wall channel
with a single smooth bump at the wall. The main physical conclusion of our work is contained in the numerical 
results plotted in Fig.~\ref{fluxcontri} and the resulting explanation of the origin of drag in terms of 
vorticity shed due to flow separation from the bump. 

It is interesting to ask how our results are related to the views of Feynman 
on the role of vortex reconnections in superfluid turbulence. Posing the question ``What can eventually 
become of the kinetic energy of the vortex lines?," \textcite{feynman1955application} argued that 
``the lines (which are under tension) may snap together and join connections a new way" and he proposed
a picture of a sequence of reconnections as a path to dissipation of vortex energy into elementary excitations.
A modern version of this picture is the Kelvin wave cascade generated by vortex 
reconnections\cite{kivotides2001kelvin,kozik2004kelvin}. In fact, the experiments of 
\textcite{bewley2008characterization} and \textcite{fonda2016sub,fonda2019reconnection} have vizualized 
the quantized vortex lines in superfluid turbulence and observed their reconnection dynamics. We agree
with the view that vortex reconnection is an essential part of turbulence, not only in quantum fluids 
but also in classical fluids. A major difference is that classical vorticity distributions are continuous 
and Newtonian viscosity allows vorticity to diffuse like smoke through the fluid. However, the stochastic 
Lagrangian description of classical vortex motion via a Feynman-Kac representation shows that 
line-reconnection occurs everywhere in classical turbulent flows, continuously in time\cite{constantin2011stochastic,eyink2020Astochastic,eyink2020Bstochastic}.
On the other hand, focusing 
on the small-scale dissipation of fluid-mechanical vortex motions into heat, in our opinion, misses an 
essential element of turbulent dissipation. Referring to classical fluid turbulence driven by a pressure
gradient, \textcite{feynman1955application} argued that ``The vortex lines twist about in an ever more 
complex fashion, increasing their length at the expense of the kinetic energy of the main stream.''
In fact, complex, irregular motion is not sufficient to explain turbulent dissipation in such flows. The essential
new idea supplied by \textcite{Josephson65} and \textcite{Anderson66}, which was missed by Feynman, 
is that organized {\it cross-stream vortex motion} and not just random stretching and reconnection
is required to explain the enhanced energy dissipation in wall-bounded turbulence of both quantum and classical fluids. 

In our opinion, this point is likely of key importance in the explanation of the {\it anomalous energy dissipation}
for incompressible fluid turbulence, which was proposed by \textcite{onsager1949statistical}$^{,}$\cite{eyink2006onsager,eyink2024onsager} and which was the subject of pioneering empirical investigations by 
\textcite{sreenivasan1984scaling,sreenivasan1998update} and \textcite{meneveau1991multifractal}. 
Various experiments\cite{nikuradse1933laws,cadot1997energy} have shown that the presence of wall-roughness 
is crucial for the existence of a dissipative anomaly and some phenomenological scaling theories\cite{gioia2006turbulent,goldenfeld2006roughness} lead to the same conclusion. Experimental
visualizations of flow around individual cubic roughness elements in a turbulent duct flow\cite{gao2021experimental}
exhibit similar features as our smooth bump, with form drag, flow separation and vortex shedding into the interior.
It thus seems likely that such phenomena must persist in order to produce a dissipative anomaly in the 
infinite Reynolds number limit. It is known from experimental studies of \textcite{sreenivasan1987unified} and \textcite{sreenivasan1997persistent} that viscous effects persist 
in the log-layer of smooth-wall turbulent flows up to the location of peak Reynolds stress and mean vorticity flux
in particular is dominated by viscous transport over this range\cite{eyink2008,kumar2023flux}. Mathematical analysis \cite{quan2022inertial,quan2022onsager}$^{,}$\cite{eyink2024onsager} shows that anomalous viscous 
transport of vorticity outward from the wall may in fact persist in the infinite-Reynolds limit, and 
persistent shedding of vorticity and resultant form drag seem the most plausible mechanism 
for anomalous energy dissipation in rough-walled turbulent flows. 

In future work, we hope to apply our new detailed Josephson-Anderson relation to several problems
of current interest. Our work gives a new perspective on the problem of turbulent drag reduction which we 
plan to pursue, in particular for polymer additives\cite{kumar2024josephson}. Note that the polymer stress
contributes simply a body force $\mathbf{f}=\bm{\nabla\cdot\tau}_p$ in the Navier-Stokes equation 
\eqref{eq_mom} and the detailed JA-relation hence applies directly to viscoelastic fluids. Another problem
of practical importance is the parameterization of surface drag in rough-walled turbulent flows, which has 
already been investigated\cite{aghaei_2022} by the Force Partition Method (FPM)\cite{menon_mittal_2021a,menon_mittal_2021b} 
which is closely related to the Josephson-Anderson relation. The relationship of these two approaches deserves 
to be discussed at length, but we just note here that FPM derives an exact expression for form drag as 
a spatial integral of the second-order invariant $Q=-(1/2){\rm Tr}\,[(\bm{\nabla}\mathbf{u})^2]$ and the 
viscous acceleration $\nu \Delta\mathbf{u}$ weighted by a scalar potential $\phi$ and its gradient 
$\bm{\nabla}\phi,$ respectively. While such an integral relation is similar in form to the JA relation, 
FPM uses a different potential, yields results for the pressure contribution to drag only, and has the aim to 
relate form drag to $Q$-structures rather than to vorticity dynamics. Another approach to derive exact 
formulas for skin friction is that of \textcite{fukagata2002contribution}, yielding the so-called FIK identity, 
and a vorticity-based version in particular relates the skin friction to velocity-vorticity correlations\cite{yoon2016contribution}, similar to the JA-relation. However, FIK-type identities apply only to 
flat-walled flows without form drag and yield a result only for homogeneous averages. The detailed Josephson-Anderson 
relation derived by \textcite{Huggins1970a} and extended in this work, by contrast, describes the total drag 
from both skin friction and form drag and applies instantaneously in time. 

\acknowledgments 
We thank Y. Du, N. Goldenfeld, J. Katz, C. Meneveau, R. Mittal and T. Zaki for discussions of this problem and of their related results. We wish to express our gratitude to K.~R. Sreenivasan for his friendship over many years and for 
his leadership in science, which we hope will continue well into the future. Finally, we thank the Simons Foundation for support of this work through the Targeted Grant No. MPS-663054, ``Revisiting the Turbulence Problem Using Statistical Mechanics'' and also the Collaboration Grant No. MPS-1151713, ``Wave Turbulence''.

\bibliography{aipsamp}

\end{document}


\preprint{AIP/123-QED}

\title{Supplementary Materials for ``A Josephson-Anderson relation for drag in classical channel flows with streamwise periodicity: 
Effects of wall roughness''}
\author{Samvit Kumar}
\author{Gregory L. Eyink}%
 \email{skumar67@jh.edu, eyink@jhu.edu}
\affiliation{ 
Department of Applied Mathematics and Statistics, Johns Hopkins University}%

\date{\today}
\maketitle

\tableofcontents

\newpage 

\section{Huggins' Reference Potential Flow}

\noindent 
The Fig.3 in the main text plotted the wall-normal components of Huggin's reference potential flow 
velocity at the inflow and outflow cross-sections. Here we plot the other two velocity components.

\begin{figure*}[h!] 
\makebox[\textwidth][c]{\includegraphics[width=0.8\textwidth]{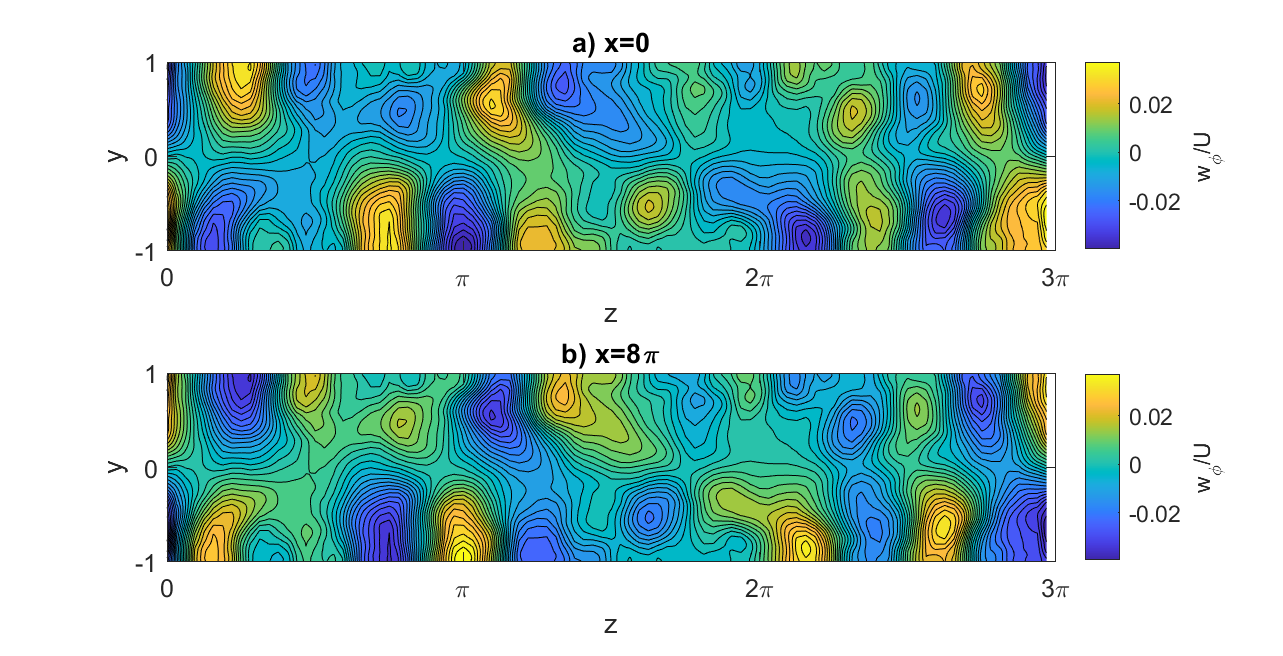}} 
\caption{
Anti-periodic spanwise component of Huggins' potential flow velocity for a flat-wall turbulent channel flow,
at inflow $x=0$ and outflow $x=8\pi.$}
\label{wphi_in_out.fig}\end{figure*}

\noindent
The above plot of the spanwise component of Huggin's reference potential velocity shows that it, 
like the wall-normal component, is streamwise anti-periodic. 

\begin{figure*}[h!]
\makebox[\textwidth][c]{\includegraphics[width=1.3\textwidth]{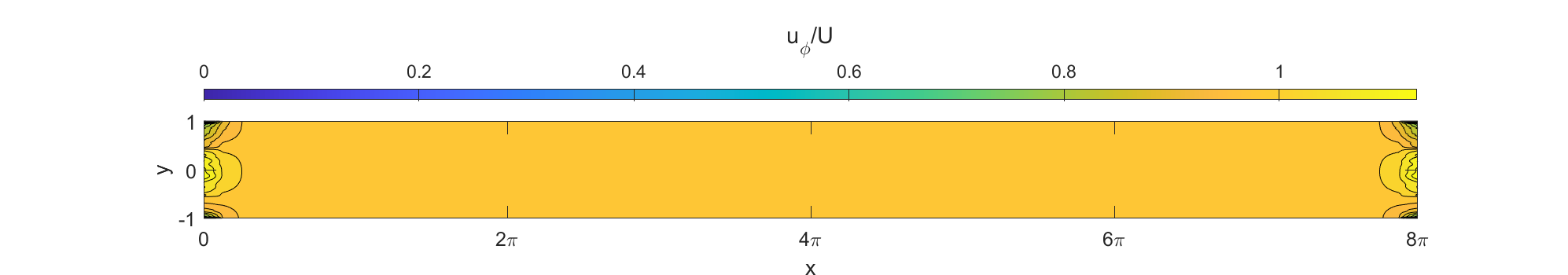}}
\caption{
Periodic streamwise component of Huggins' potential flow velocity for a flat-wall turbulent channel flow, 
in the spanwise midplane $z=3\pi/2$. }
\label{uphichannel}
\end{figure*}

\noindent
The streamwise component of the reference potential velocity in Huggin's construction is guaranteed to be 
periodic, as verified in the above plot. Inertial adjustment toward plug flow away from 
inflow/outflow is observed, just as in the plot of the streamlines in Fig.~2a of the main text. 

\clearpage

\section{Velocity and Vorticity Fields at Drag Maximum \& Minimum}

\noindent 
Plotted here are the same fields in the same flow sections as appear in Fig.~8 of the main text
for the time $tU/L_x=9.75,$ but now at times of local maximum and minimum drag.

\begin{figure}[h!]
      \centering
     \begin{subfigure}[b]{0.32\textwidth}
         \centering
         \includegraphics[width=\textwidth]{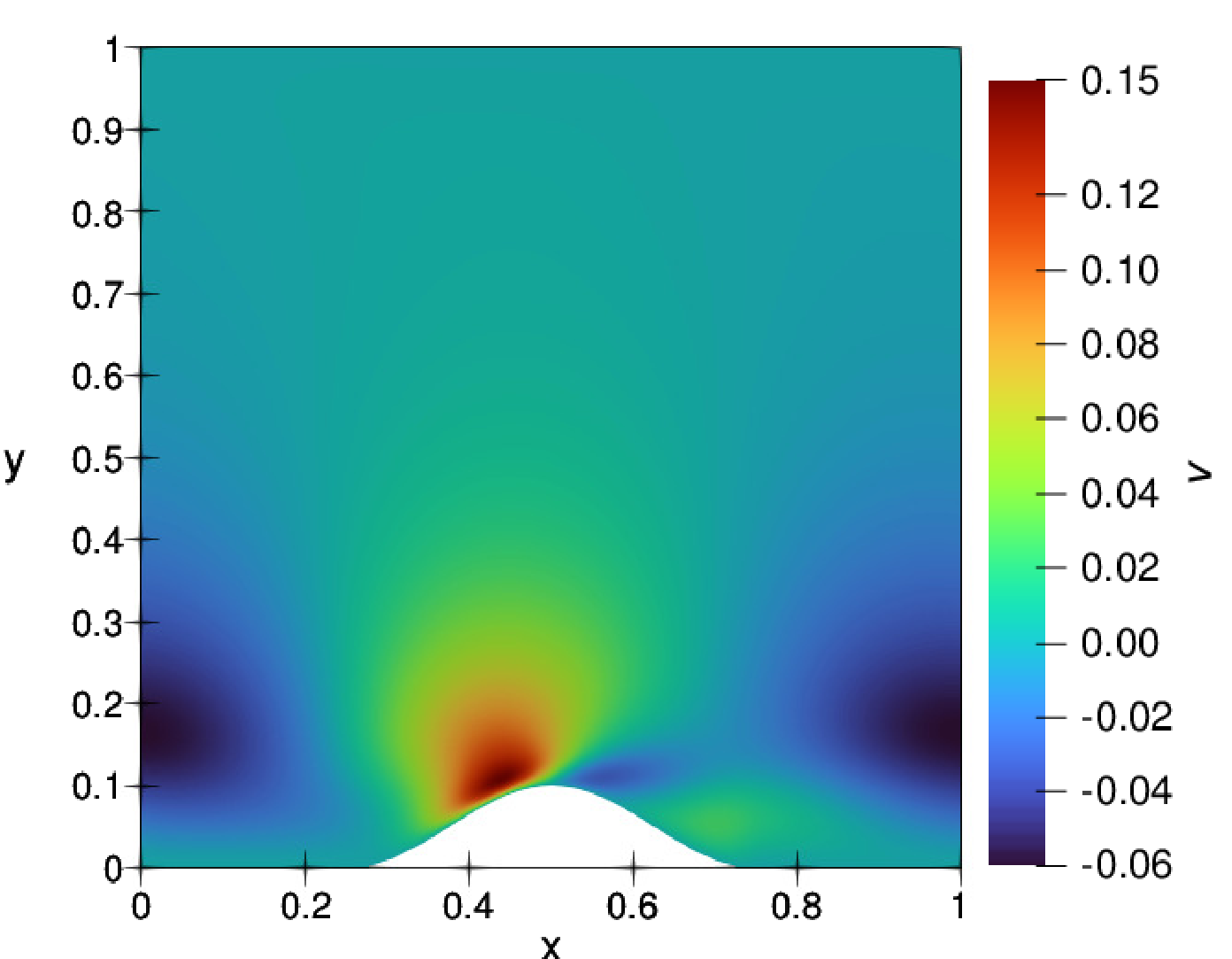}
        \caption{Wall-normal velocity }
         \label{vinst3p6}
     \end{subfigure}
     \hfill
         \begin{subfigure}[b]{0.32\textwidth}
         \centering
         \includegraphics[width=\textwidth]{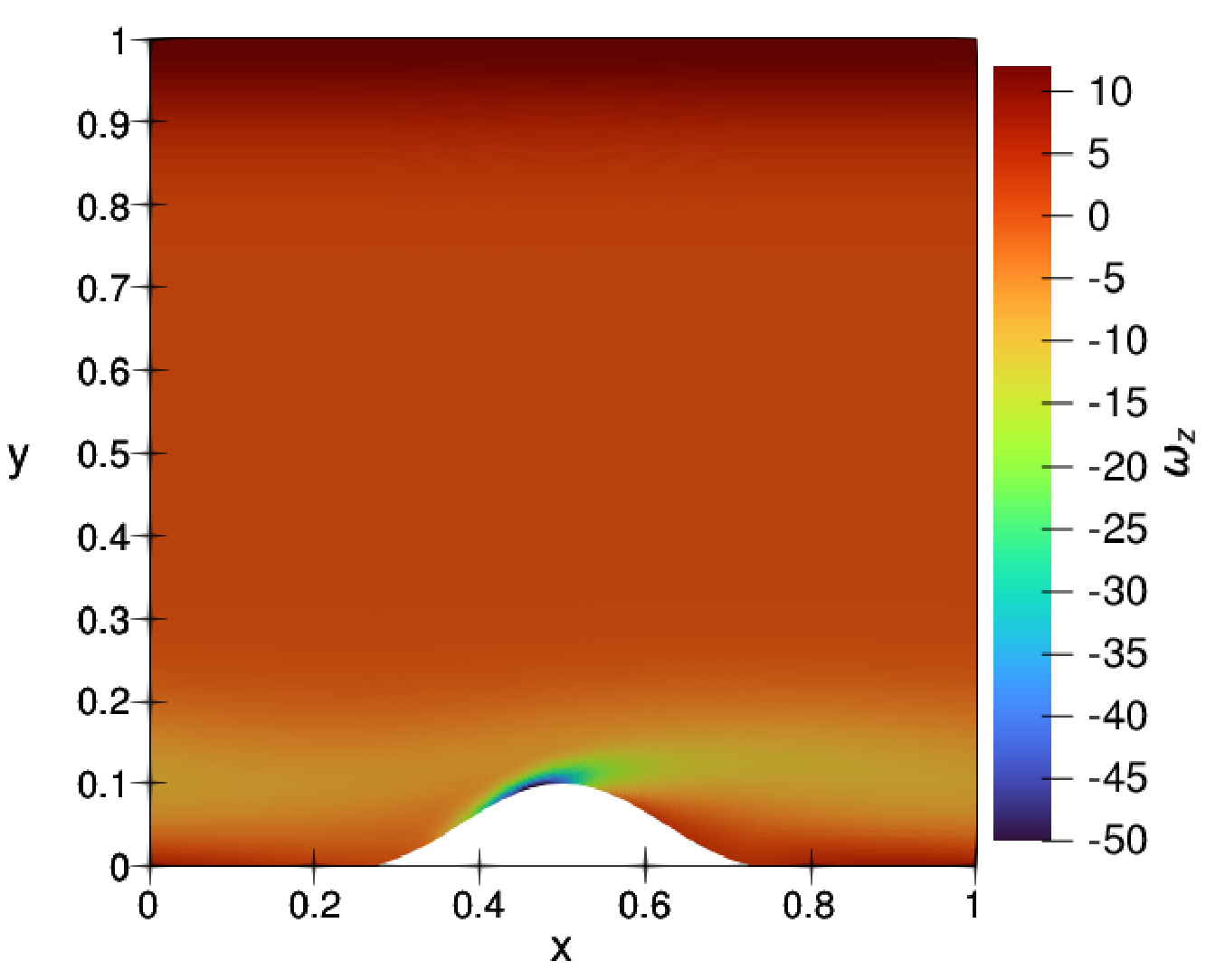}\caption{Spanwise vorticity }
         \label{ozinst3p6}
     \end{subfigure}
              \caption{Instantaneous (a) wall-normal velocity, normalized by $U$, and (b) spanwise vorticity field normalized by $U/L_x$, at spanwise plane $z=0.25$ and time $tU/L_x=3.51$ of local maximum drag}   \label{fields3p6}
\end{figure}

\begin{figure}[h!]
      \centering
     \begin{subfigure}[b]{0.32\textwidth}
         \centering
         \includegraphics[width=\textwidth]{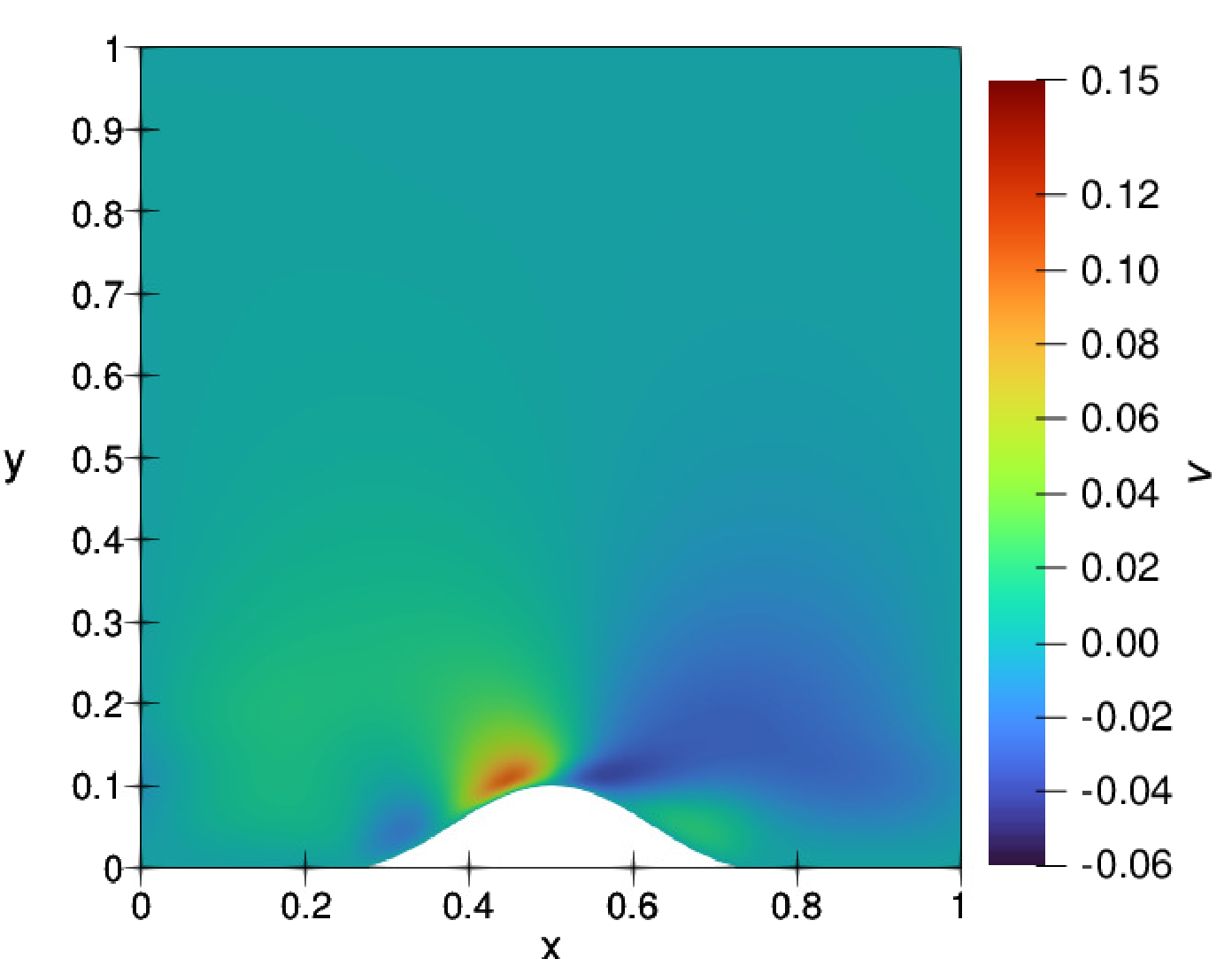}
        \caption{Wall-normal velocity }
         \label{vinst4p6}
     \end{subfigure}
     \hfill
         \begin{subfigure}[b]{0.32\textwidth}
         \centering
         \includegraphics[width=\textwidth]{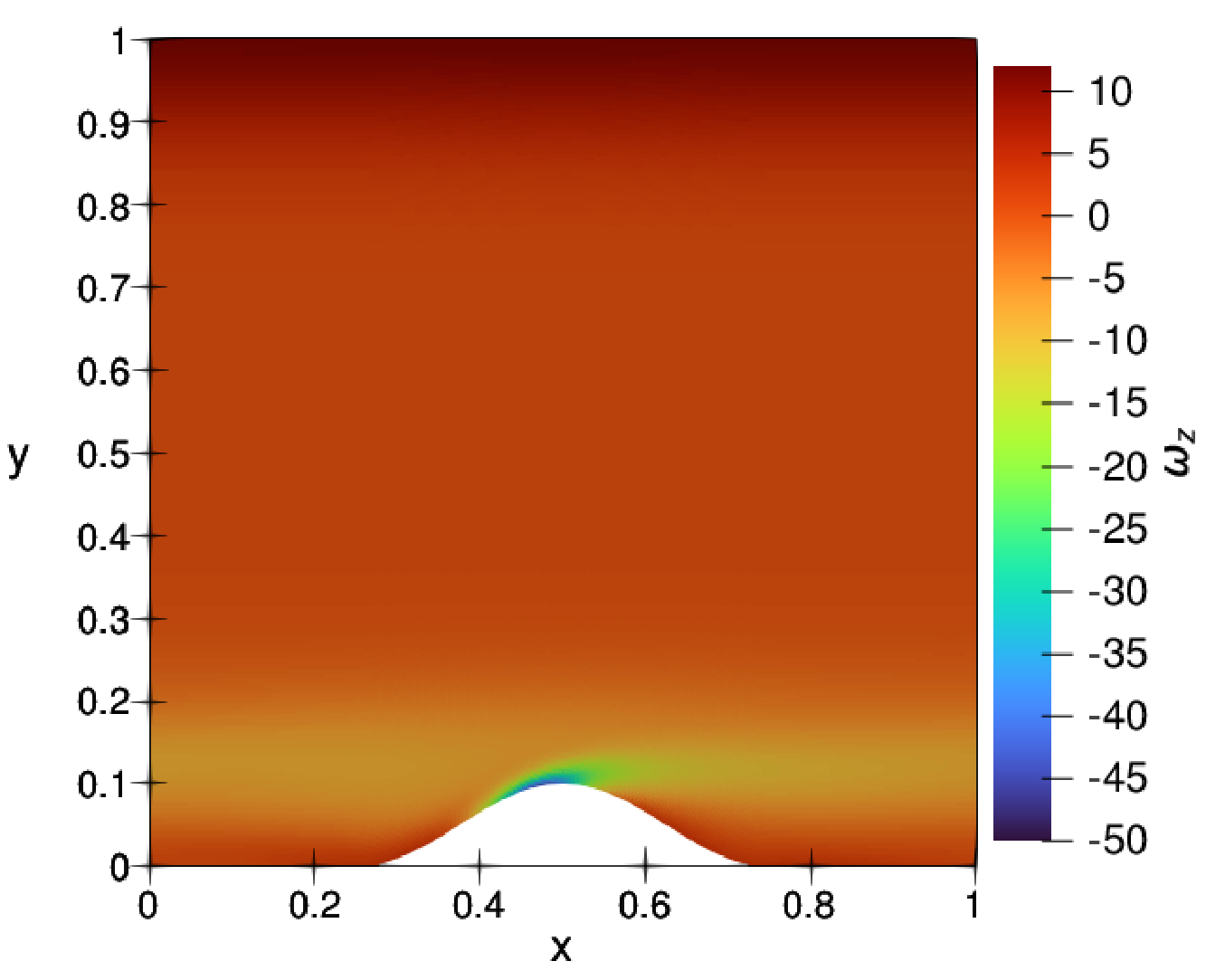}\caption{Spanwise vorticity }
         \label{ozinst4p6}
     \end{subfigure}
              \caption{Instantaneous (b) wall-normal velocity, normalized by $U$, and (b) spanwise vorticity field normalized by $U/L_x$, at spanwise plane $z=0.25$ and time $tU/L_x=4.485$ of local minimum drag}   \label{fields4p6}
\end{figure}

\noindent The main features which are apparent at the time of local maximum drag are the larger wall-normal 
velocities and also the more intense negative spanwise vorticity on the upstream side of the bump. Note that the 
point of separation on the downstream side is almost the same at both times. 

\clearpage 

\section{JA Transfer Integrands At Drag Maximum \& Minimum}

\noindent 
Plotted here are the same fields in the same flow sections as appear in Fig.~9 of the main text
for the time $tU/L_x=9.75,$ but now at times of local maximum and minimum drag.

\begin{figure}[h]
      \centering
      \begin{subfigure}[b]{0.32\textwidth}
         \centering
         \includegraphics[width=\textwidth]{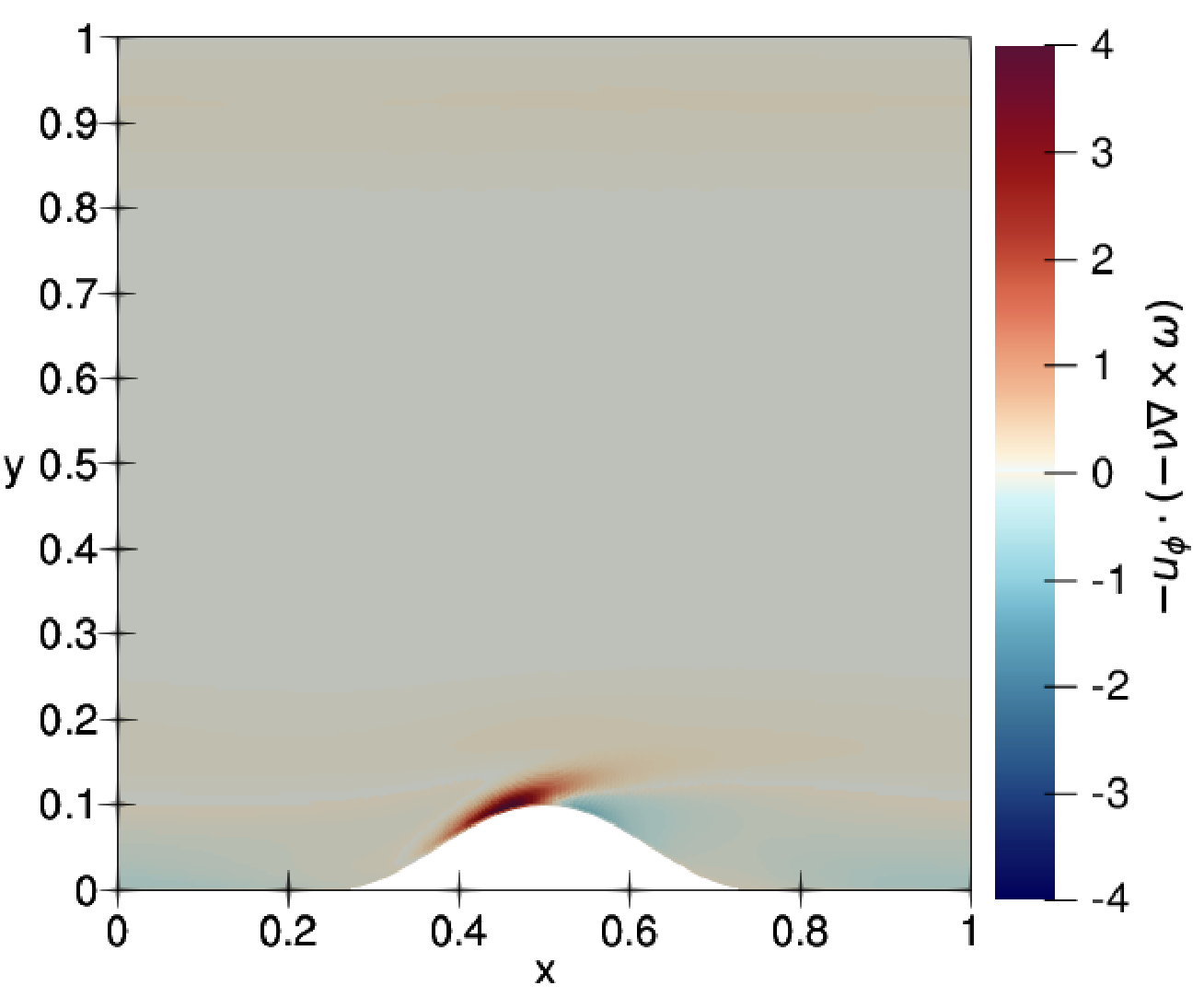}
        \caption{Instantaneous viscous contribution, giving $67.1\%$ of $\mathcal{T}$. }
         \label{visc3p6}
     \end{subfigure}
     \hfill
         \begin{subfigure}[b]{0.32\textwidth}
         \centering
         \includegraphics[width=\textwidth]{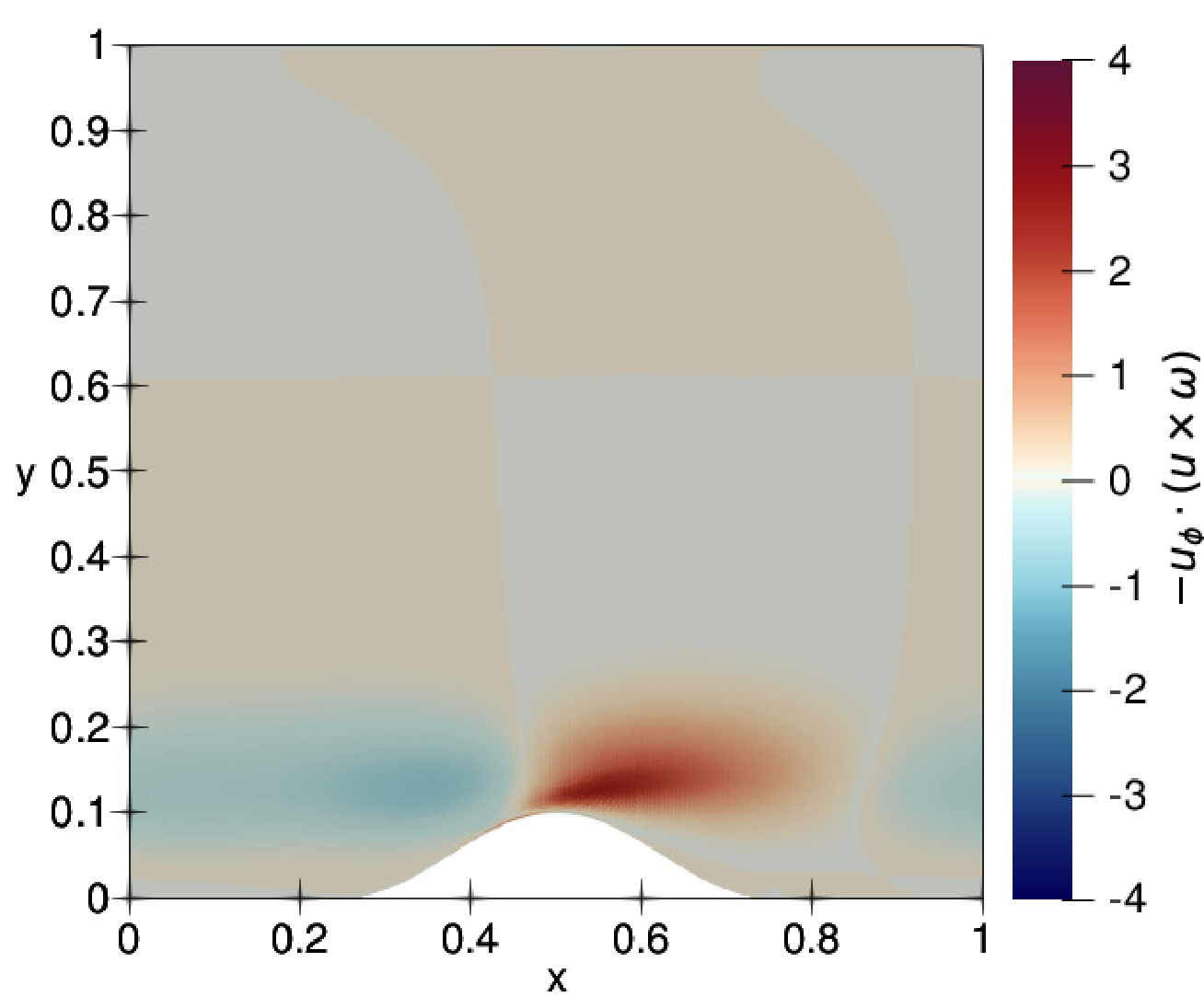}\caption{Instantaneous nonlinear contribution, 
         giving $32.9\%$ of $\mathcal{T}$. }
         \label{nl3p6}
     \end{subfigure}
     \hfill
     \begin{subfigure}[b]{0.32\textwidth}
         \centering
         \includegraphics[width=\textwidth]{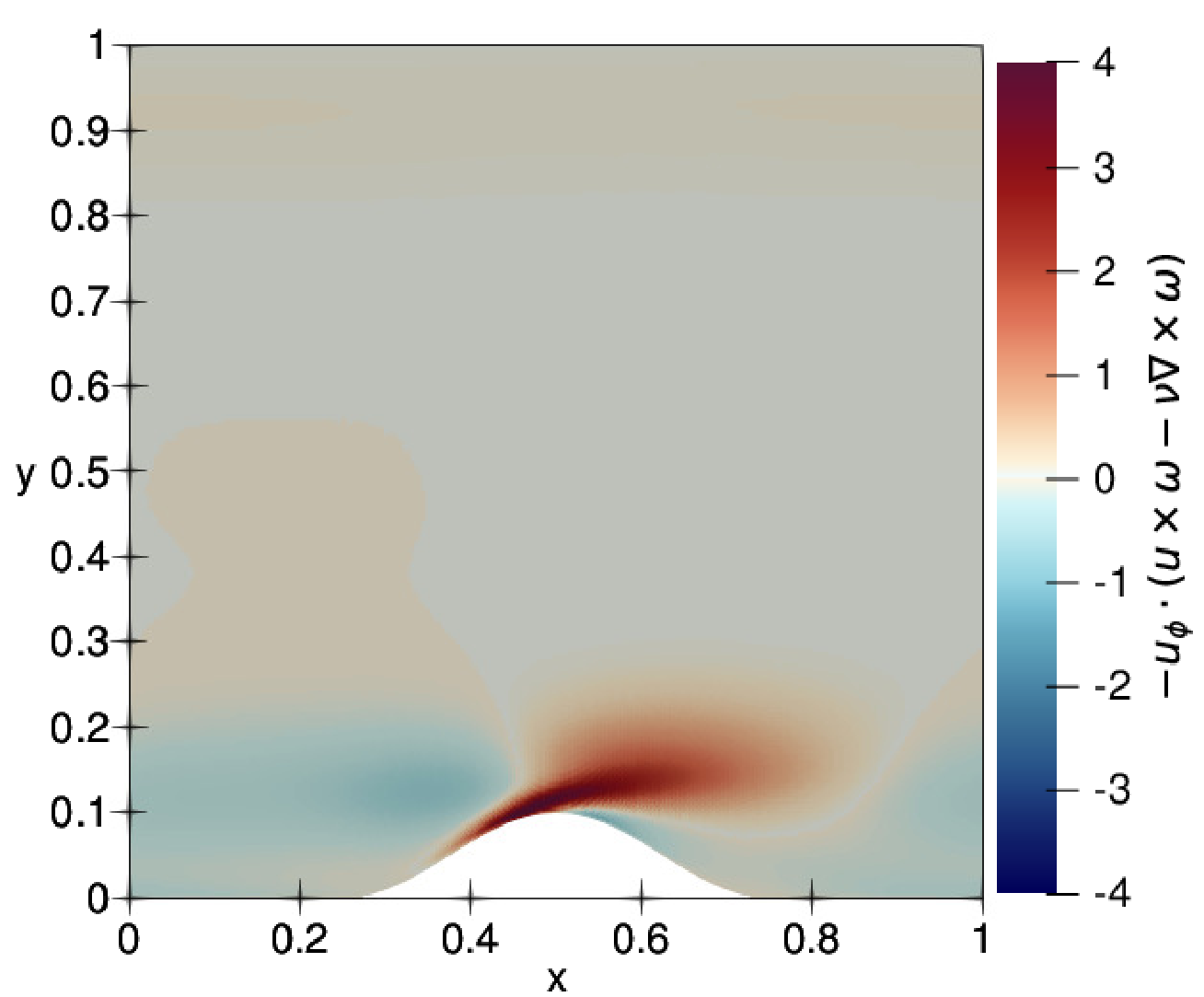}
         \caption{Instantaneous integrand of $\mathcal{T}$ in the
         JA-relation}
         \label{syz3p6}
    \end{subfigure}
                 \caption{Instantaneous fields of (a) the integrand of $\mathcal{T}$ , (b) viscous and (c) nonlinear contributions to the integrand, all normalized by $\rho U^3/L_x$. Fields are shown at spanwise plane 
                 $z=0.25$ and time $tU/L_x=3.51$ of local maximum drag}   \label{flux3p6}
\end{figure}
\begin{figure}[h]
      \centering
      \begin{subfigure}[b]{0.32\textwidth}
         \centering
         \includegraphics[width=\textwidth]{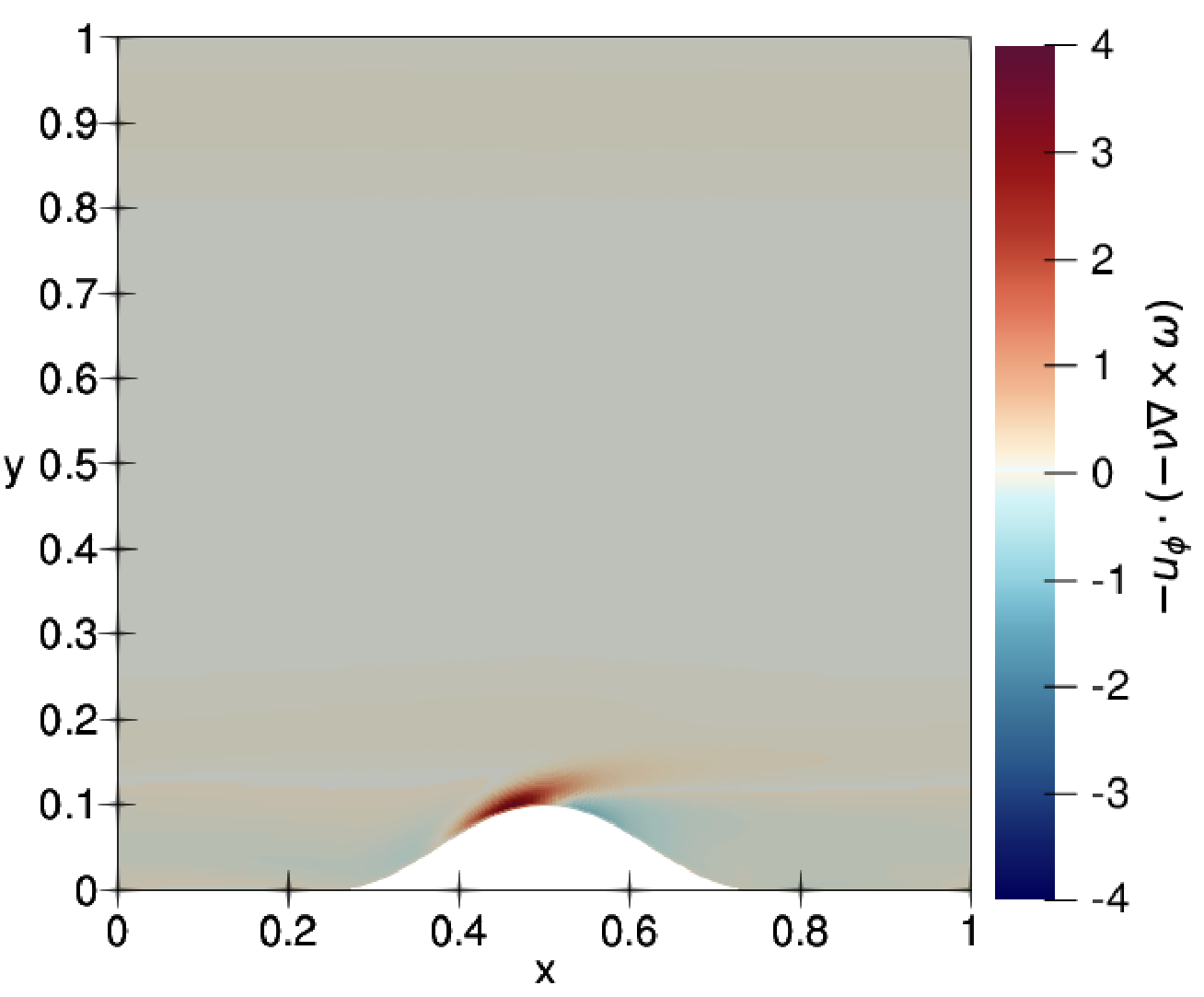}
        \caption{Instantaneous viscous contribution, giving $83\%$ of $\mathcal{T}$. }
         \label{visc4p6}
     \end{subfigure}
     \hfill
         \begin{subfigure}[b]{0.32\textwidth}
         \centering
         \includegraphics[width=\textwidth]{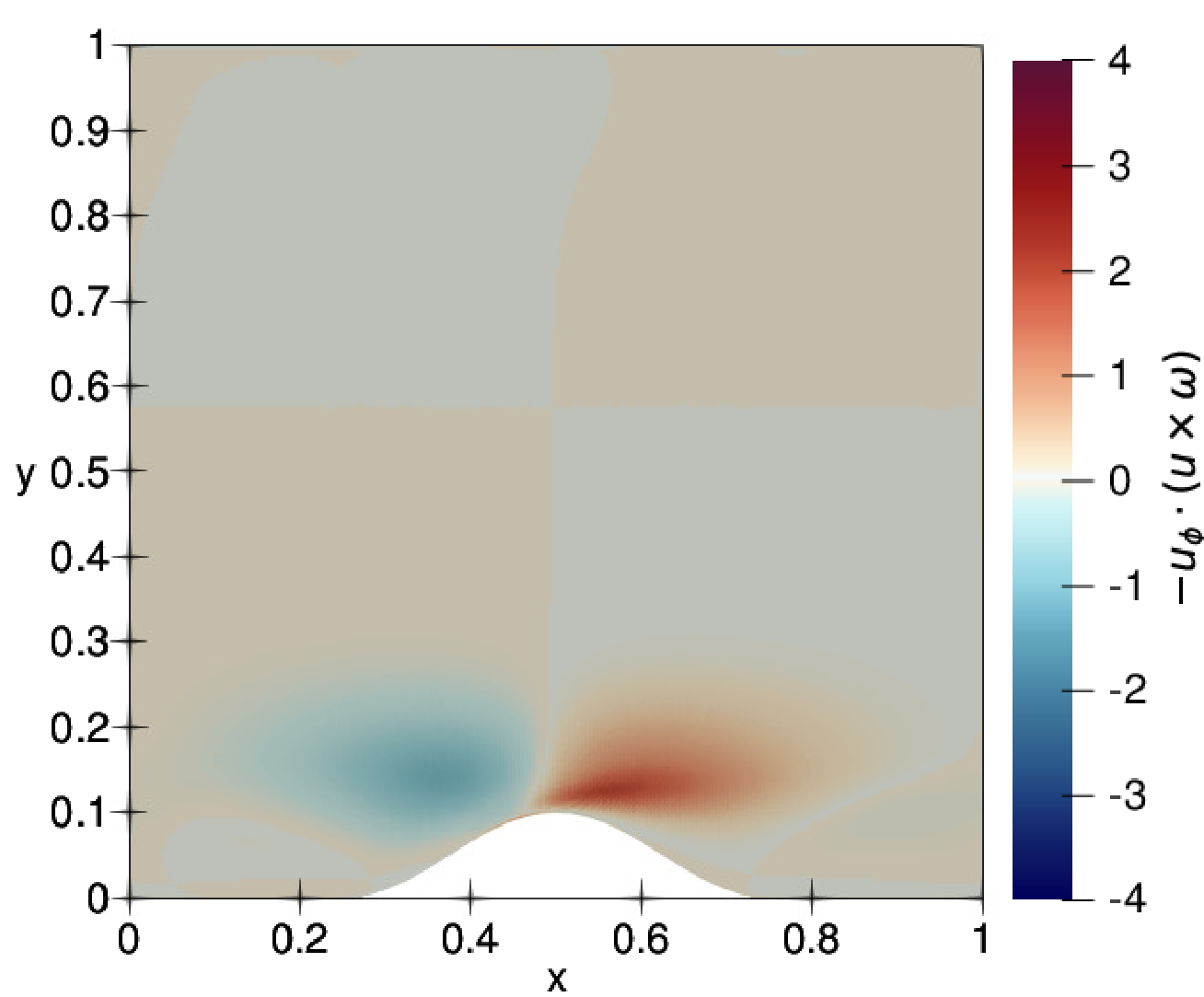}\caption{Instantaneous nonlinear contribution, giving $17\%$ of $\mathcal{T}$. }
         \label{nl4p6}
     \end{subfigure}
     \hfill
     \begin{subfigure}[b]{0.32\textwidth}
         \centering
         \includegraphics[width=\textwidth]{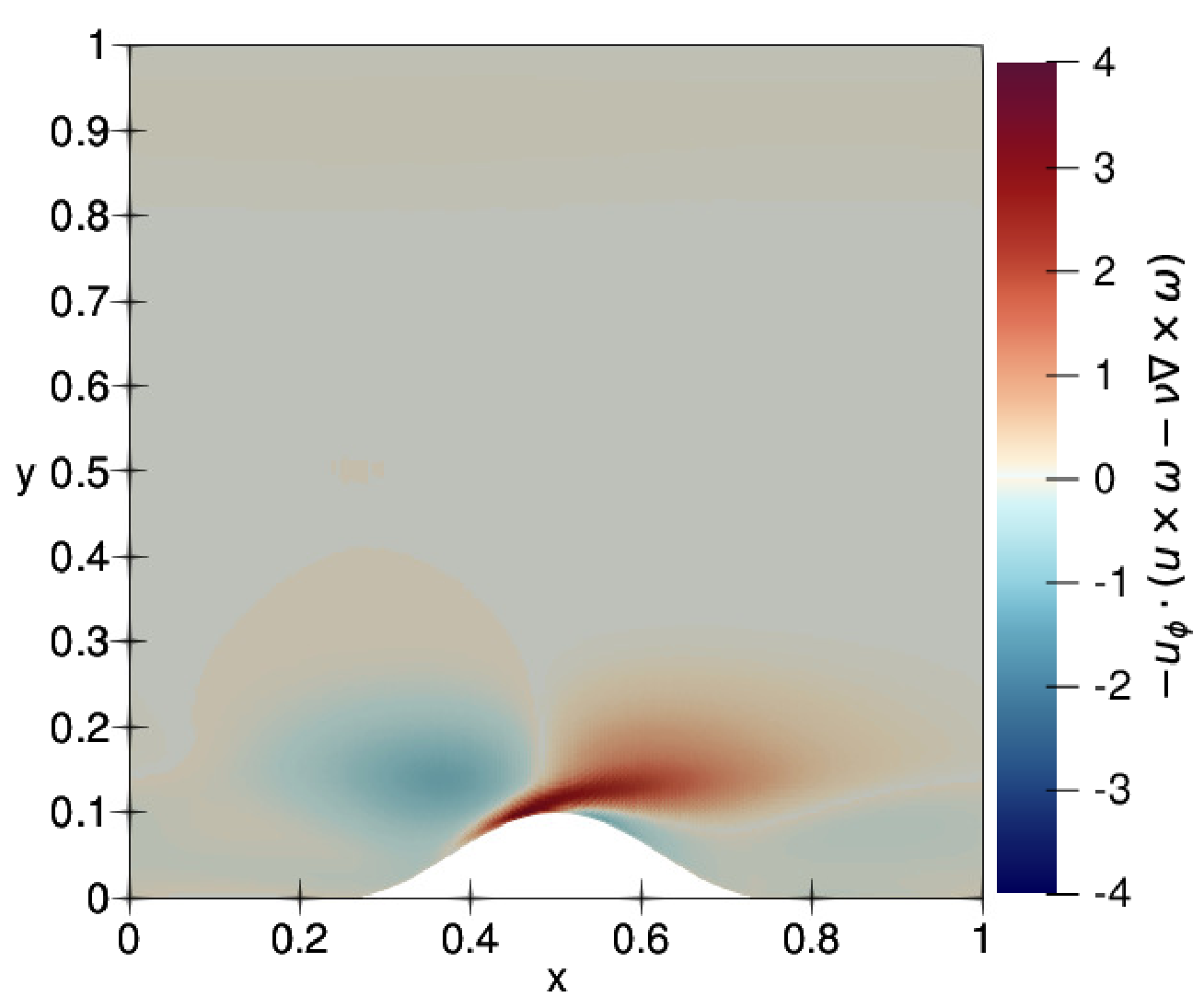}
         \caption{Instantaneous integrand of $\mathcal{T}$ in the JA-relation. }
         \label{syz4p6}
    \end{subfigure}
                 \caption{Instantaneous fields of (a) the integrand of $\mathcal{T}$ , (b) viscous and (c) nonlinear contributions to the integrand, all normalized by $\rho U^3/L_x$. Fields are shown at spanwise plane 
                 $z=0.25$ and time $tU/L_x=4.485$ of local minimum drag}   \label{flux4p6}
\end{figure}

\noindent A main result is the stronger contribution from vortex shedding at the time of local maximum drag. Also 
notable is a somewhat stronger "anti-drag" lobe upstream at the time of local minimum drag. 

\clearpage

\section{Pressure Fields}

We show here the pressure fields at several key instants in our flow. 

\begin{figure}[!h]
      \centering
      \begin{subfigure}[b]{0.45\textwidth}
         \centering
         \includegraphics[width=\textwidth]{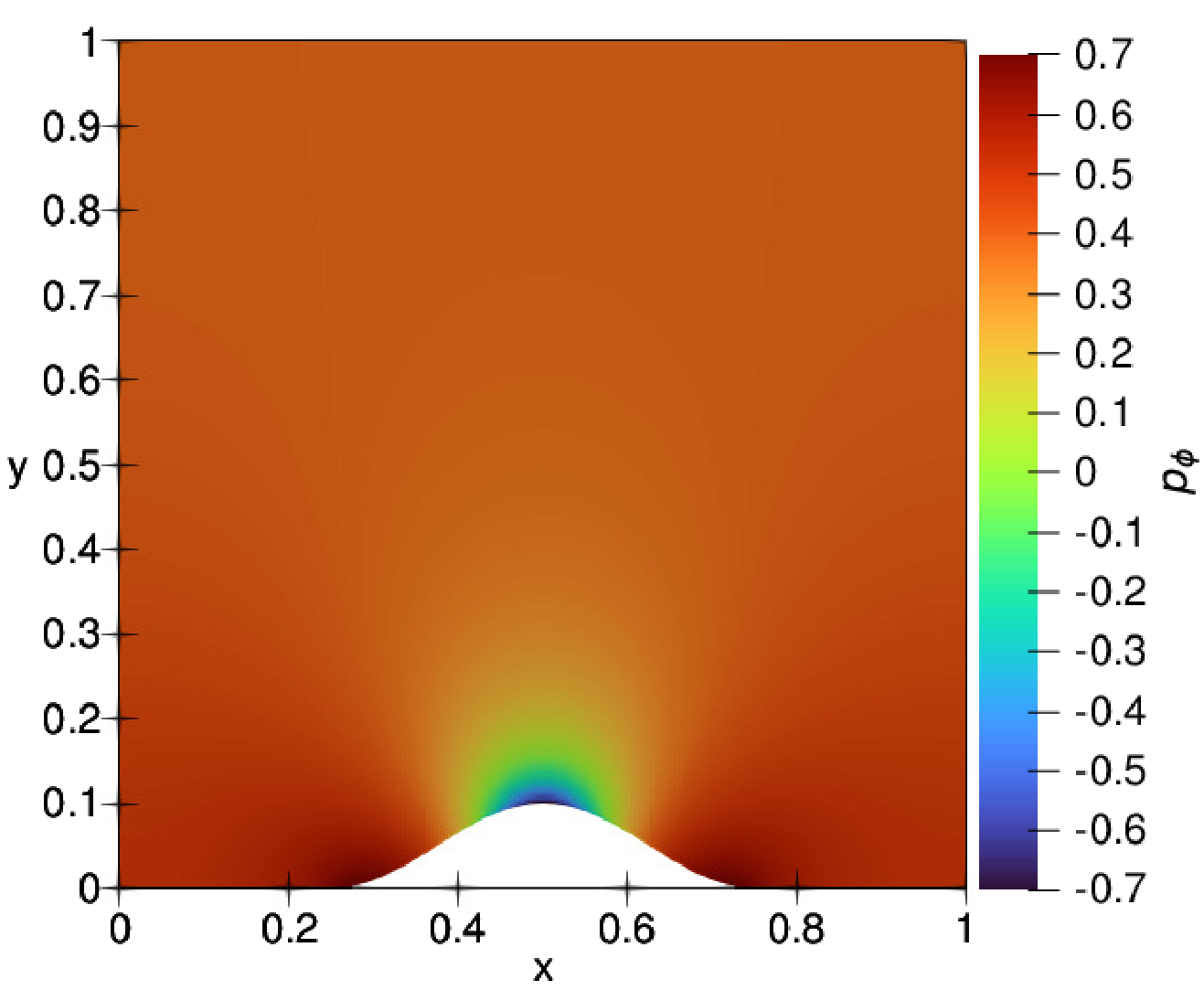}
        \caption{Pressure $p_{\phi}$ of reference flow}
         \label{pphi}
     \end{subfigure}
     \hfill
         \begin{subfigure}[b]{0.50\textwidth}
         \centering
         \includegraphics[width=\textwidth]{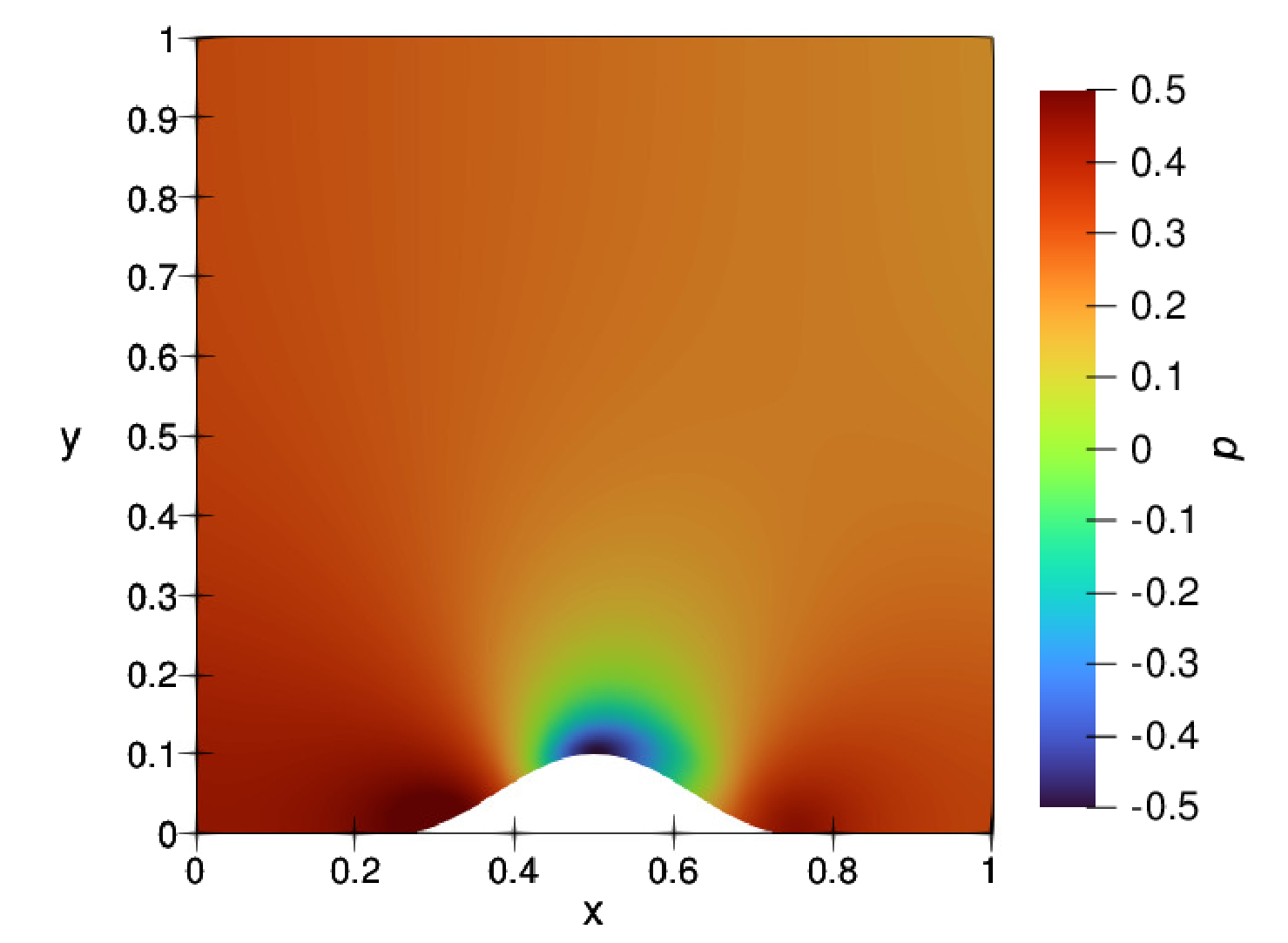}
         \caption{Pressure $p$ at time $tU/L_x=0.195$} 
         \label{pres0p2}
     \end{subfigure}
 
        \caption{Pressure of the background potential Euler flow and pressure at an early time}    \label{pres_init}
\end{figure}


\noindent 
The pressure $p_\phi$ of our new reference potential Euler solution can be calculated from the Bernoulli relation, $p_{\phi}+(1/2)\rho |\mathbf{u}_{\phi}|^2=c(t)$. For the flow over the bump described in the main text of the paper, the associated potential pressure field is shown in Fig.~\ref{pphi}.  Naturally, there is no form drag for the current case, where the bulk velocity $U$ is constant in time, and the potential pressure is perfectly symmetric about $x=0.5.$ The Navier-Stokes 
flow discussed in the main text is initialized with plug flow, which is close to $\mathbf{u}_\phi,$ and the pressure 
$p$ from the Poisson equation at early times is close to $p_\phi,$ as illustrated in Fig.~\ref{pres0p2} for time $tU/L_x=0.195.$
However, a slight asymmetry has developed in $p$ due to rotational pressure $p_\omega,$ which results in some form drag.  

\clearpage

\begin{figure}[!h]
      \centering
      \begin{subfigure}[b]{0.32\textwidth}
         \centering
         \includegraphics[width=\textwidth]{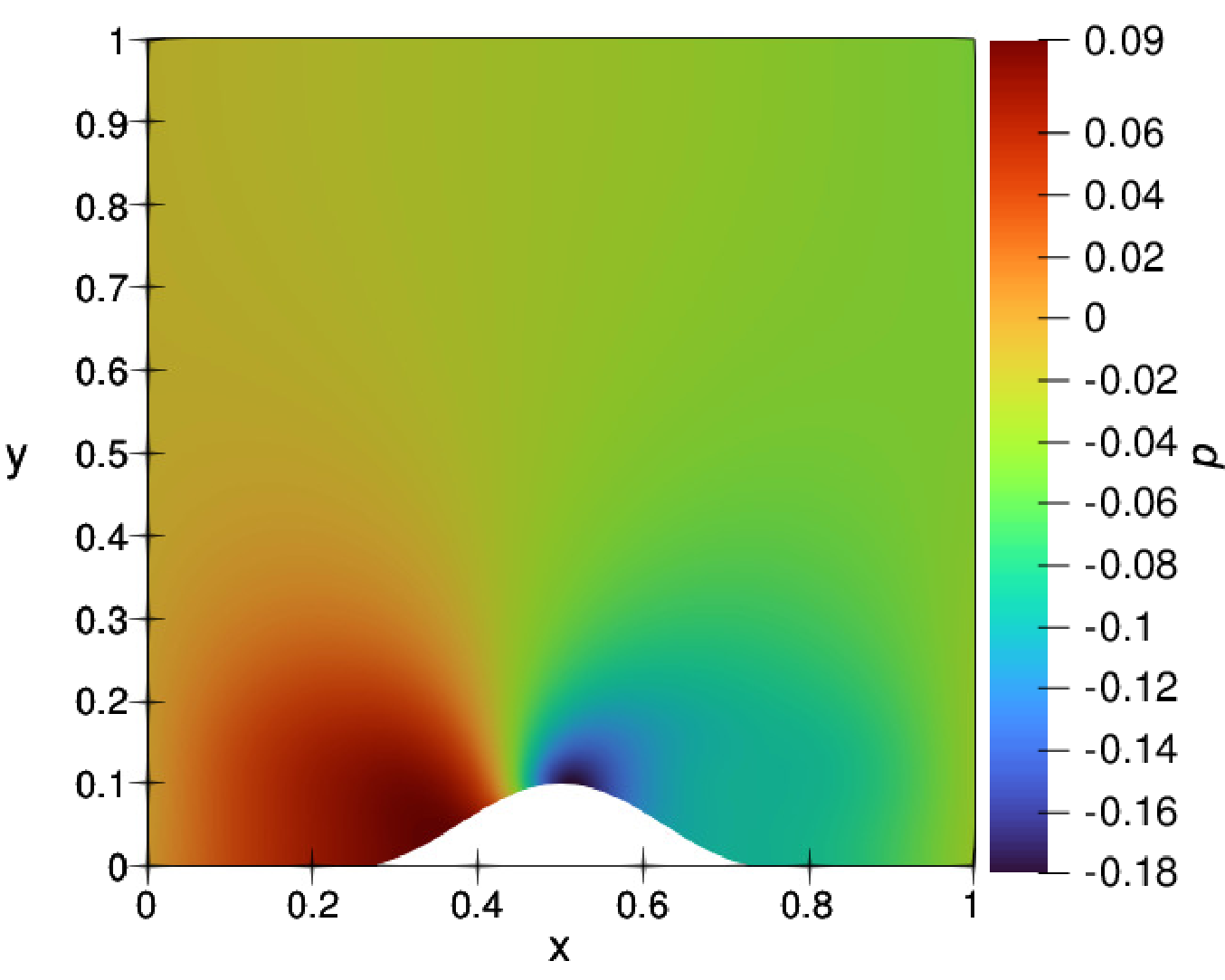}
        \caption{}
         \label{pres3p6}
     \end{subfigure}
     \hfill
         \begin{subfigure}[b]{0.32\textwidth}
         \centering
         \includegraphics[width=\textwidth]{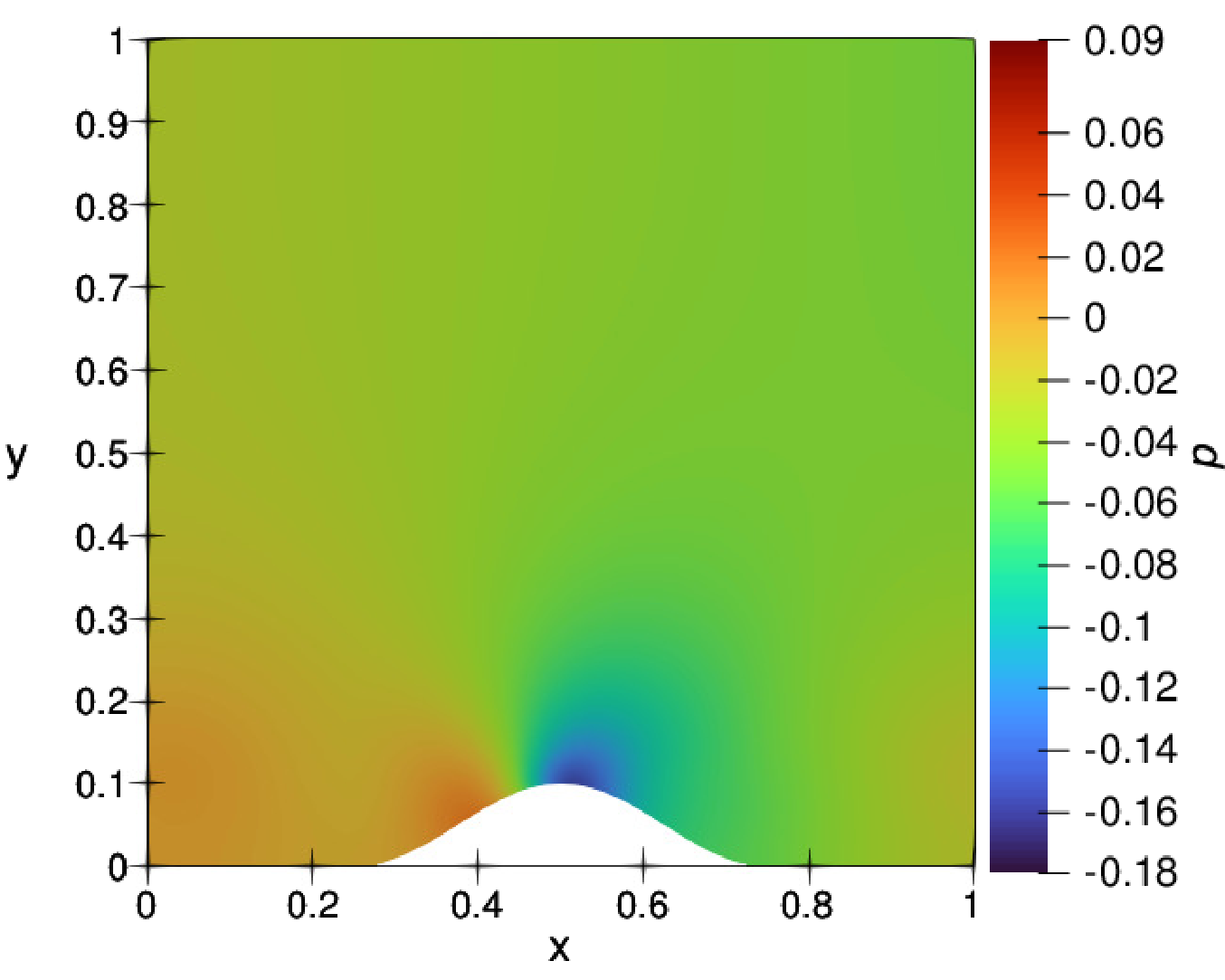}\caption{}
         \label{pres4p6}
     \end{subfigure}
     \hfill
     \begin{subfigure}[b]{0.32\textwidth}
         \centering
         \includegraphics[width=\textwidth]{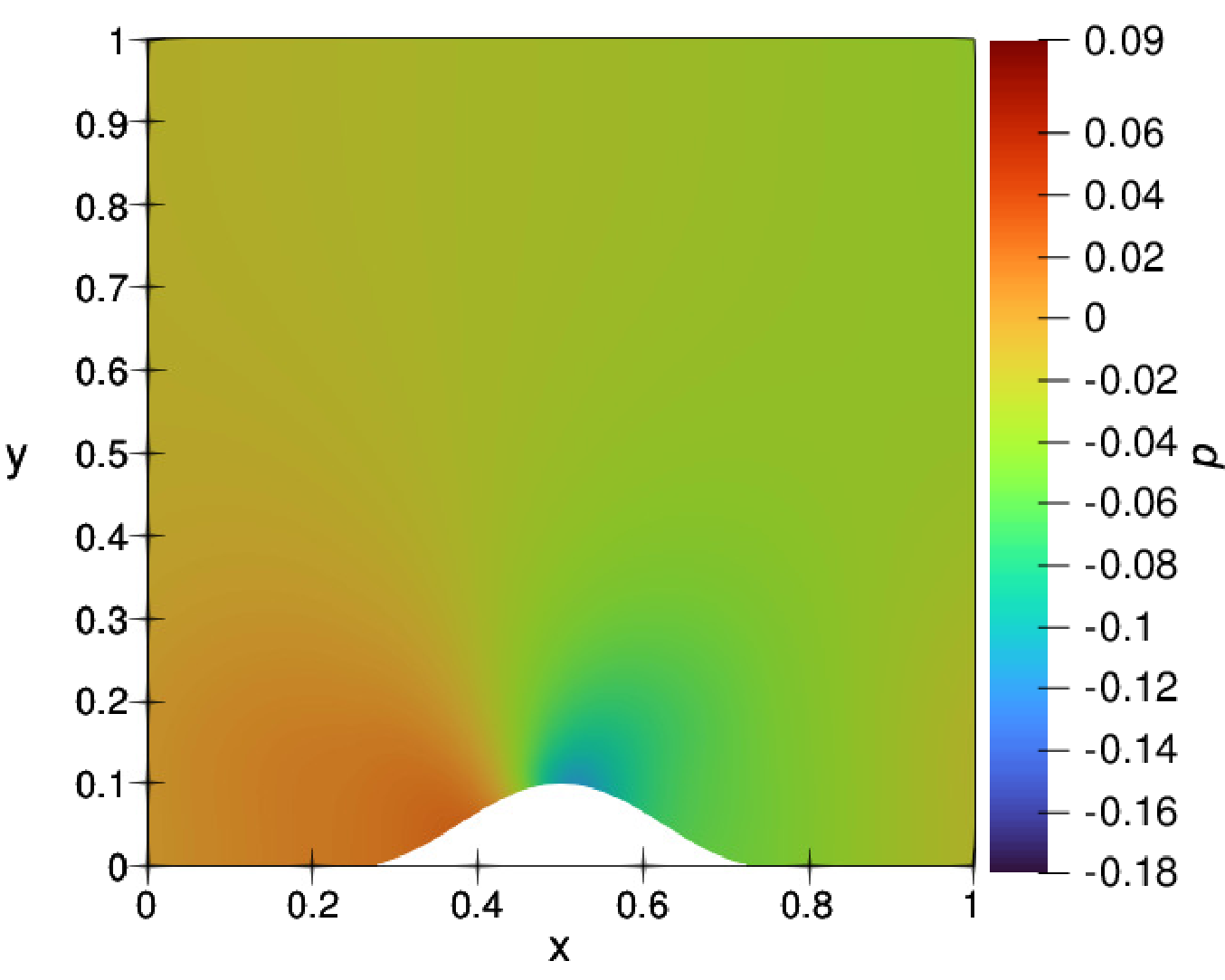}
         \caption{ }
         \label{pres10}
    \end{subfigure}
                 \caption{Instantaneous pressure fields at z=0.25 for (a) $ tU/L_x=3.51$, (b) $tU/L_x=4.485$ (c) $tU/L_x=9.75$, all normalized by $\rho U^2$. }   \label{fluxcontri}
\end{figure}

\noindent 
We show instantaneous pressure fields at a time of local maximum drag ($tU/L_x=3.51$) in Fig.~\ref{pres3p6}, 
at a time of local minimum drag ($tU/L_x=4.458$) in Fig.~\ref{pres4p6}, and at the late time $tU/L_x=9.75$
discussed in detail in the main text in Fig.~\ref{pres10}. The pressure difference on the upstream and downstream 
faces of the bump is much higher for the local drag maximum when compared with the other two instants. 
This is consistent by Lighthill's relation $\bm{\sigma}=-\mathbf{n}\bm{\times\nabla}p$ with the greater magnitude 
of negative spanwise vorticity shed into the flow in this case, 
as observed in Fig.~\ref{ozinst3p6}. Notice that the pressure magnitudes have dropped significantly at all 
of these times compared to the potential pressure $p_\phi$ and the pressure $p$ at the early time $tU/L_x=0.195$ 
presented in Fig.~\ref{pres_init}. 

\clearpage
